\documentclass[12pt,journal,draftcls,a4paper,onecolumn]{IEEEtran}
\IEEEoverridecommandlockouts

\usepackage[utf8]{inputenc} 
\usepackage[T1]{fontenc}
\usepackage{cite}
\usepackage{float}
\usepackage[belowskip=-8pt,aboveskip=2pt]{caption}
\usepackage{textcomp}

\ifCLASSINFOpdf
  \usepackage[pdftex]{graphicx}

\else

\fi

\usepackage[cmex10]{amsmath}

\usepackage{multirow}
\usepackage{array}
\usepackage[lofdepth,lotdepth]{subfig}
\usepackage[pdftex,bookmarks=true]{hyperref}
\hypersetup{
  colorlinks   = true, 
  urlcolor     = black, 
  linkcolor    = black, 
  citecolor   = blue 
}

\usepackage{multirow}
\usepackage{array}
\usepackage[lofdepth,lotdepth]{subfig}
\usepackage{authblk}
\usepackage{amssymb}
\renewcommand{\thetable}{\arabic{table}}
\def\tablename{Table}

\begin{document}


\title{Relay Selection with Imperfect SIC for FD/HD NOMA Cooperative Networks over Nakagami-$m$ Fades*}  


\author{Efendi Fidan, Eray Güven,  Mehmet Akif Durmaz,  Güneş Karabulut Kurt, and O\u{g}uz Kucur
\thanks{*This work is supported by the Scientific and Technological Research Council of Turkey (TUBITAK) under Grant EEEAG 118E274 and will be submitted to IEEE Journals.}
\thanks{E. Fidan is with the TUBITAK National Metrology Institute (TUBITAK UME), Gebze-Kocaeli, Turkey, 41470, (e-mail: efendi.fidan@tubitak.gov.tr).}
\thanks{O. Kucur is with the Department of Electronics Engineering, Gebze Technical University, Gebze/Kocaeli, 41400, Turkey (e-mail: okucur@gtu.edu.tr).}
\thanks{E. Güney, M. A. Durmaz,  and G. K. Kurt are with the Department of Electronics and Communication Engineering, Istanbul Technical University, Istanbul, Turkey, 34469, (e-mail: \{durmazm, guvenera, gkurt\}@itu.edu.tr).}
}

\maketitle
\begin{abstract}

In this work, a non-orthogonal multiple access (NOMA) based transmission between two sources and two end-users is examined over independent non-identically distributed (i.n.i.d.) slow  Nakagami-$m$ fading channels, where a single relay using decode-and-forward (DF) protocol is  selected out of a set of full-duplex/half-duplex (FD/HD) multiple relays in accordance with the quality of service criterion. Two relay selection (RS) strategies, selecting a relay to maximize data rate of  user 1 at selected relay and selecting a relay out of a set of relays providing service quality for user 1 to maximize data rate of user 2, are analyzed.   Additionally, not only perfect successive interference cancellation (pSIC) but also imperfect SIC (ipSIC) is considered.  The exact and asymptotic outage probability (OP) expressions are derived and validated via Monte Carlo simulation technique. Unlike existing works, our expressions are unique and valid for all cases such as FD and HD together with  pSIC and ipSIC, i.e. expressions are not given separately but in a single compact form.  Effect of each component such as pSIC, ipSIC, and self-interference (SI) for FD transmission on error floor of OP is demonstrated. Moreover, the optimum  relay location is illustrated for a plenty of scenarios consisting of combination of different power allocations, data rates, pSIC/ipSIC, and total transmitted powers.

\end{abstract}

\begin{IEEEkeywords}
NOMA, Relay Selection, Full-Duplex, Half-Duplex, Decode-and-Forward, Nakagami-$m$, Imperfect Successive Interference Cancellation .
\end{IEEEkeywords}

\newcounter{MYtempeqncnt}
\IEEEpeerreviewmaketitle
\IEEEpubidadjcol
\section{Introduction}

Effective resource utilization of the communication resources such as latency, power and spectrum are essential and critical issues that should be handled  carefully in wireless networks. A promising technique to  satisfy massive connectivity, low latency, and  high spectral efficiency for the fifth generation (5G) and beyond networks is non-orthogonal multiple access (NOMA), which also achieves higher user fairness \cite{ding2017survey, aldababsa2018tutorial, ding2015cooperative, lv2018cognitive, liu2016fairness,shin2017non}. In NOMA, all resource blocks, i.e., non-orthogonal resources can be used by each user; therefore, as compared to traditional orthogonal multiple access (OMA) techniques where available resource blocks are occupied separately,  spectrum efficiency is improved \cite{ding2015cooperative, dai2015non}. NOMA can be implemented in different domains such as code and power domains or mixed of them \cite{vaezi2019multiple}, where superposed signals can be simultaneously received (up-link NOMA) or transmitted (down-link NOMA). In code domain,  a unique spreading code is assigned to each user, similarly, a portion of the total power is assigned to each user in power domain. Successive interference cancellation (SIC) is used to separate the superposed signals. However, due to the limited processing power at the end-users, power domain NOMA seems to be the dominant choice and gain more attention in the literature: Two different power allocation (PA) strategies are available, namely, assigning more power to the users having poor channel conditions or allocating power in accordance with  quality of service (QoS) requirements, where PA factors may or may not be independent of channel state information (CSI) or not. Considering CSI during PA means always assigning more power to the user facing worse fading, on the other hand, more power is assigned to a user with QoS priority where CSI is not considered. 

Cooperative NOMA is first introduced in \cite{ding2015cooperative} for a system consisting of multiple-users and a base station (BS) over Rayleigh fading channels, where less power is allocated to the users with better channel conditions. Each user applies SIC and decodes its signal and thereafter, they share decoded signals via short range communication channels to avoid the use of extra time slots. Since cooperative relaying networks (CRN) extend coverage area, consume less power, and improve reliability and security; their application on NOMA-based transmission systems over many fading environment and forwarding protocols, namely, amplify/decode-and-forward (AF/DF) protocols with both half-duplex (HD) and full-duplex (FD) relays has been  deeply investigated in terms of ergodic capacity and outage probability (OP) with adaptive and fixed PAs \cite{kim2015non, men2016performance, ding2016relay, yang2017novel,  deng2017joint, yue2018spatially, zhao2018dual, xu2018optimal, chen2018performance, alkhawatrah2019buffer, li2019joint, liu2019performance, tang2019performance, ju2019performance, lee2016non, Junior2019performance, li2019performance, lei2019secrecy, yu2019secrecy, wang2019secrecy, feng2018two, lei2019physical}. Contrary to OMA-based systems \cite{krikidis2012full,efendi2019Performance, fidan2021performance}, opportunistic relay selection (RS) is not possible for NOMA-based systems due to the fact that at least two users are served and a single selected relay can not enhance reliability for all users. Therefore, other approaches such as sub-optimum \cite{xu2018optimal, ju2019performance, tang2019performance}, buffer-aided \cite{alkhawatrah2019buffer}, partial \cite{lee2016non, Junior2019performance},  and multi-level (two-stage) \cite{ding2016relay, yang2017novel, yue2018spatially,chen2018performance} RS schemes are offered and analyzed.   Sub-optimum selection strategies are based on conventional max-min criterion,  partial relay selection (PRS) strategies are based on selecting a relay with the maximum channel gain between the source (BS) and relays, two stage RS (TSRS) schemes are based on selecting a relay from  a predetermined subset of relays realizing an assigned condition such as priority of a user or sum of users' data rates, which has also been proposed for OMA-based physical layer security \cite{zou2015relay, zhu2016security, zou2016relay, guo2019joint, guo2017joint}. RS in case of buffer-aided transmission is done in accordance with a number of predetermined conditions where not only relay link between the BS and relay providing target data rate but also transmission mode (NOMA, time-division-multiple-access) is also selected, which only implements downlink NOMA in the second hop (between the selected relay and end-users)\cite{alkhawatrah2019buffer}. Additionally, TSRS scheme with the  condition that at least $N$ relays should be in the predetermined subset is known as the $N^{th}$ best RS scheme \cite{liu2019performance}. Without loss of generality, hereinafter, we use the term CRN to emphasize a cooperative relaying network consisting of  a BS, multiple-relays, and two end-users (destinations).

In \cite{ju2019performance}, a sub-optimum RS based on max-min criterion for a CRN with HD relays is offered over non-selective Rayleigh fade but the implementation is similar to FD relaying operation   except that the selected relay is not used in the subsequent transmission but a relay is selected from the remaining set, where all the relays keep receiving transmitted signals of the BS during broadcasting time slot of the selected relay. Therefore, self (loop)-interference (SI) effect is eliminated but inter-relay interference due to previously selected relay is not taken into account and it is not possible to cancel out this interference as in FD relaying. By implementing the fixed PA policy and giving priority to the far user, OP and ergodic sum rate are analyzed. The proposed selection method outperforms existing HD RS schemes but due to FD operation, comparisons are not fair. A comprehensive comparison with FD RS schemes will complete the analysis of proposed RS method.  Another, sub-optimum RS using max-min strategy is investigated for a two-way transmission system consisting of two pairs of NOMA users exchanging data via multiple relays by implementing the downlink and uplink NOMA protocols \cite{tang2019performance}. OP performance is illustrated together with hardware impairments for fixed power factors and slow Rayleigh fading environment, where DF relays are used.

Conventional max-min (sub-optimum) RS and TSRS schemes are introduced and compared for a CRN over slow Rayleigh fading channels to select an HD DF relay in \cite{ding2016relay}. In this work, end-users are not ordered according to their CSI but rather categorized by QoS requirements, i.e., fixed PA is used and priority is given to user 1. Firstly an HD subset of relays guaranteeing user 1's target data rate at all nodes (relay, user 1, and user 2) is determined and thereafter, a relay maximizing data rate of user 2 is determined. OP is analyzed and it is proven that offered TSRS scheme outperforms the conventional max-min RS scheme. A similar system is considered in \cite{yang2017novel} and a slightly different condition is assigned to determine the subset of HD AF/DF relays: The set of relays realizing data rate of both users at relays and user 1 are determined, i.e., subset of relays satisfies QoS requirements of user 1 is detected  and then, the best relay is selected out of this subset to serve user 2, where realization of data rate of user 1 at user 2 is not ensured. The results demonstrate that offered selection criterion has better performance as compared to previous works. In \cite{xu2018optimal}, two TSRS strategies termed as two-stage weighted--max--min (WMM) and max--weighted--harmonic--mean (MWHM), are proposed for cooperative NOMA with fixed and adaptive PAs at the relays, respectively. Since NOMA is only applied in the second hop, in the first hop a coded message including data of both users is transmitted so no need of SIC at the relay, relay-subset is determined based on the relays achieving sum data rate of two users. The results demonstrate that the proposed RS schemes outperform the existing TSRS schemes. OP of another different TSRS strategy is studied in \cite{li2019performance}, where PAs are dynamically determined in accordance with the poorness of CSI, i.e., more power is allocated to the poor links between the relay and end-users. Relay subset is determined by taking the sum of data rates as a threshold, where a relay is in the selected subset if its capacity is larger than or equal to sum of the data rates but additionally, QoS priority for user 1 is also ensured. Similar to the work in \cite{xu2018optimal}, the considered CRN has multiple HD DF relays and the BS  transmitting a superposed signal produced by Gray mapping based joint-modulation \cite{yan2015receiver} so that both users’ messages can be decoded simultaneously without the complicated SIC. Besides, such a coding strategy also leads to lower OP than superposition coding  at the relays.  Furthermore, nearly the same TSRS scheme is simultaneously  investigated in spatially random CRN over non-selective block Rayleigh fading with fixed PAs and QoS requirement for user 1 and the same condition is applied to determine relay subset in \cite{yue2018spatially, chen2018performance} except that work in  \cite{yue2018spatially}  considers DF HD/FD relays but that of \cite{chen2018performance} focuses on DF/AF HD relays. A critical issue that should be pointed out is the dilemma of zero/non-zero diversity order between these two works; non-zero diversity order for HD transmission is claimed in the former one but zero-diversity order is in the later one. In the context of this work, the reason for that dilemma is also clarified. Finally, physical-layer security based on NOMA protocol is also deeply examined as OMA counterpart based on TSRS schemes in \cite{lei2019secrecy, yu2019secrecy, wang2019secrecy, feng2018two, lei2019physical}. 
 
In this work, two different relay selection schemes, namely, single-stage relay selection (SSRS) and TSRS strategies in accordance with the service quality priority are applied to a system consisting of two sources and two users (destinations) over independent non-identically distributed (i.n.i.d.) slow  Nakagami-$m$  fading channels. A single relay using DF is selected out of a set of FD/HD multiple relays in accordance with the quality of service criterion, where first user is assumed to have service priority. Both  perfect SIC (pSIC) and imperfect SIC (ipSIC) situations are considered. For practical approaches, it is reasonable and more comprehensive to investigate such systems over Nakagami-$m$ fading channel since it covers many fading environments such as Rayleigh and Rician. Moreover, pSIC is difficult to achieve in practical applications, i.e., not possible in reality, therefore, ipSIC should be analyzed to shed light on performance of  such communication systems. A system with a single source is investigated in \cite{fidan2021performance} with similar conditions over i.i.d. Generalized-$K$ fading channels.

\subsection{Contributions}

The main contributions of this work can be summarized as follows:

\begin{itemize}
	\item  Exact and asymptotic OP expressions for both RS strategies, namely, SSRS for the user 1 and TSRS for the user 2 are derived over  i.n.i.d. slow  Nakagami-$m$  fading channels with not only pSIC but also ipSIC, and their validity is verified via Monte Carlo simulation technique.  Unlike existing works, our expressions are unique and valid for all cases such as FD and HD together with  pSIC and ipSIC, i.e., expressions are not given separately but in a single compact form.
	
	\item   At the high signal-to-noise (SNR), asymptotic expression  of SSRS explicitly reveals that throughput is almost not affected by PA factors at the selected relay and characteristics (shaping and scaling factors) of  the channels among the selected relay and the two end-users. However, it is dominated by PA factors at the sources and characteristics  of the channels among the sources and selected relay and that of SI channel for FD transmission. For TSRS scheme, previously mentioned deductions are also valid for pSIC case but characteristics of  the channel between the selected relay and user 2 and  ipSIC at user 2 become effective on throughput when ipSIC takes place. Additionally, characteristics of ipSIC at the selected relay are also other substantial  factors dominating throughput at the high SNR.

	\item Optimum relay location for many scenarios consisting of different PA factors, data rates, SIs for FD transmission,  total transmitted powers, and ipSIC at the relays and user 2 are handled out and illustrated. The results demonstrate that relay location changes toward to the second source for FD transmission to eliminate SI effect and toward to the first source for HD transmission to eliminate SIC effect.

	\item Modeling of SI variance  as $\kappa_l P_{R_i}^{\vartheta_l-1}$, where $\kappa_l$ and $\vartheta_l$ are linear and exponential cancellation/attenuation factors depicting the dependence of SI on the transmitted power, $P_{R_i}$, from each relay and both vary between $0$ and $1$, allows to individually investigate effect of SIC, SI, and ipSIC on error floors.
	
	\item   At high SNR regions, signal-to-interference-plus-noise ratios (SINRs) approach constant values, in turn, error floors, i.e., equal probabilities occur for the communication systems using NOMA or FD protocol, where SI power is only linearly dependent on transmitted power. Therefore, dividing logarithm of the error (OP or symbol error rate (SER))  by logarithm of transmitted power or SNR and then letting power or SNR approach to infinity causes limit of this ratio to be zero, which is commonly accepted as zero-diversity order. Actually, treating the slope of $\log-\log$ plot as diversity order is a common myth based on the work of Wang and Giannakis, where even no diversity technique is used, slope of $\log-\log$ plot is considered as diversity order for Nakagami-$m$ fading \cite [{eq. (6)}]{wang2003simple}. The diversity order is the number of independent paths between the source and destination, i.e., number of independent scaled copies of a transmitted signal arriving to the destination but not the slope of the $\log-\log$ plot which is just only valid over Rayleigh fading channels for HD/FD transmission over Rayleigh fading channels, where SI variance dependence on transmitted power is assumed to be exponentially. But this approach is not suitable to determine diversity order of HD analysis of this work. To be more precise, a robust and comprehensive solution has already been offered but does not attract attention in recent researches, especially on those of FD and NOMA-based transmission systems\cite[eq. (1)]{godavarti2002diversity}, where diversity order is determined by comparing probabilities for the system with a diversity technique and that without any diversity technique.  It is worth noting that diversity order and independent number of channels are not always corresponding to each other\cite{smith2013number}. Lastly, non-zero diversity order notion is also elaborated in detail in  \cite{fidan2021performance}.

\end{itemize}

\subsection{Paper Outline}

The remainder of this paper is organized as follows. In Section \ref{SystemModel}, the details of system model are provided. RS criteria and derivations of exact and asymptotic OP are given in Section \ref{OutageProbability}. Numerical results are detailed in  Section \ref{NumericalResults}. Finally, we conclude our work in Section \ref{Conclusion}.


\section{System Model} \label{SystemModel} 

\begin{figure}[htbp]  
	\centering
	\includegraphics[width=4.4in]{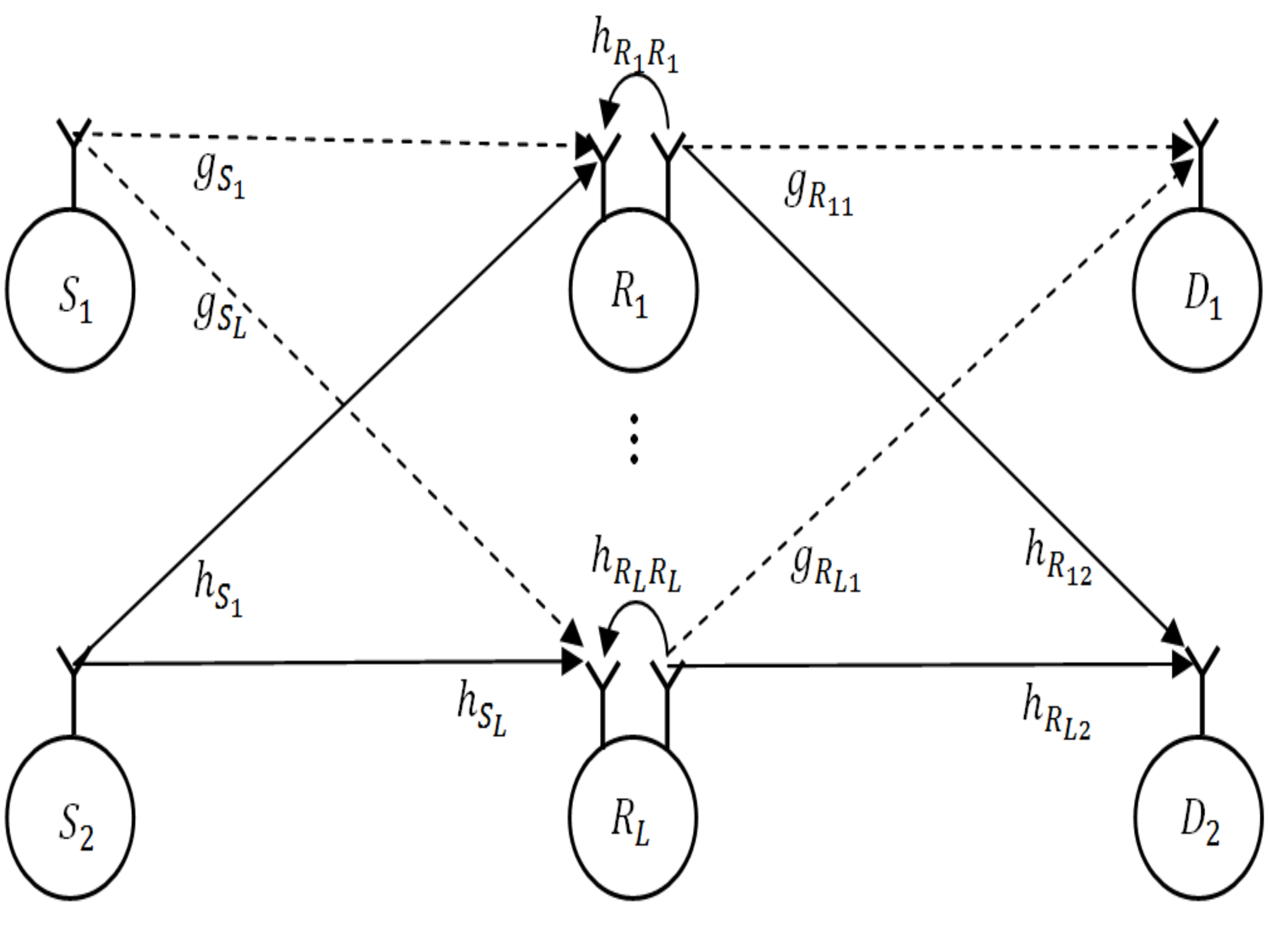}
	\caption{Multi-relay system.}
	\label{SystemDescribtion}
\end{figure}

We consider a multiple relay system with two sources and two users as depicted in Fig. \ref{SystemDescribtion}. Due to deep fading and shadowing, no direct link is assumed among the sources ($S_1$ ve $S_2$) and users ($D_1$ ve $D_2$) and the communication is conducted via $L$ FD/HD multi-relays with DF protocol. The information signals are assumed to have unit energy. A relay is selected to provide service quality priority for the user 1, $D_1$. The transmission from the sources to the selected relay and from the selected relay to the users  is done via uplink and downlink NOMA, respectively. Without loss of generality, relays are assumed to be clustered such that they are at the origin of the cartesian system and have equal distance to each node. Position of each node, the sources and users is represented as $(x_j, y_j)$, $j \in \{S_1, S_2, D_1, D_2 \} $ with respect to this origin.

To investigate  the best relay location, the assumed origin is kept fixed and positions of relays are denoted as $\bigl(x_{R_i}, y_{R_i}\bigr)=\bigl(r\cos(\theta), r\sin(\theta)\bigr)$. Whereby, distance between the node $j$ and relays  is  $d_{jR_i}=\sqrt{(x_j-r\cos(\theta))^2+(y_j-r\sin(\theta))^2}$. As in \cite{kader2017exploiting}, the sources are assumed to collaborate for PA. The PA coefficients at the $S_1$ and $S_2$ are represented as  $a_1$ and $a_2$, where $a_1+a_2=1$ and $a_1 > a_2$.  Similarly, after DF process at the selected relay, detected signals are weighted in accordance with the service quality priority by PA coefficients $a_3$ and $a_4$, where $a_3+a_4=1$ and  $a_3 > a_4$. The relays are equipped with one transmit and one receive antennas. All channel gains are assumed to have  Nakagami-$m$ distribution. Therefore, square amplitudes of channel gains are Gamma distributed. The channel gains of the links $S_1$\textemdash$i^{th}$ relay, $S_2$\textemdash$i^{th}$ relay, $i^{th}$ relay\textemdash$D_1$,  $i^{th}$ relay\textemdash$D_2$, ipSIC at the $i^{th}$ relay, ipSIC at the $D_2$, and SI are  $g_{S_i}$, $h_{S_i}$,   $g_{R_{i1}}$,  $h_{R_{i2}}$, $\tilde{g}_{S_i}$, $\tilde{h}_{R_{i2}}$ and $h_{R_iR_i}$, respectively. The shape and  scale parameters of the stated links are $(m_{S_{1i}}, \Omega_{S_{1i}}= \Omega_{S_{1i}}^0d_{S_{1}R_i}^{-\alpha})$, $(m_{S_{2i}},\Omega_{S_{2i}}= \Omega_{S_{2i}}^0d_{S_{2}R_i}^{-\alpha})$, $(m_{D_{1i}}, \Omega_{D_{1i}}=\Omega_{D_{1i}}^0d_{R_iD_{1}}^{-\alpha})$, $(m_{D_{2i}}, \Omega_{D_{2i}}=\Omega_{D_{2i}}^0d_{R_iD_{2}}^{-\alpha})$, $(\tilde{m}_{R_{i}}, 1)$, $(\tilde{m}_{D_{2i}}, 1)$, and $(m_{R_iR_i}, \Omega_{R_iR_i}\kappa_l P_{R_i}^{\vartheta_l-1})$, respectively, where $\alpha$ is path loss exponent. Terms with  superscript $0$ in the definition of scale parameters represent values where distances among nodes are not considered or taken to be 1. Total power at the sources and the $i^{th}$ relay  are   $P_S$ and  $P_{R_i}$, respectively.  

The received signal at the $i^{th}$ relay, by omitting time indexing for the sake of the simplicity, is
 \begin{equation} \label{Denklem_1}
 Y_{R_i}=\sqrt{a_1P_S}g_{S_i}x_1+\sqrt{a_2P_S}h_{S_i}x_2+\varpi_i h_{{R_i}{R_i}}x_{R_i}+n_{R_i},
 \end{equation}
where $x_{R_i}=\sqrt{a_3 P_{R_i}}\hat{x}_1+\sqrt{a_4 P_{R_i}}\hat{x}_2$,  $\hat{x}_1$ and $\hat{x}_2$ are previously decoded signals by SIC process.  $\varpi_i$ is  $0$ for HD and $1$ for FD transmission. $n_{R_i}$ is the additive Gaussian noise at the relay $i$. Decoded signals are weighted in accordance with the service quality priority and retransmitted. SINR related to the signal of the first source at the $i^{th}$ relay is 
  \begin{equation} \label{Denklem_2}
  \gamma_{D_1}^{ R_i}=\frac{a_1 \rho_{S_{i}}|g_{S_i}|^2}{a_2\rho_{S_{i}}|h_{S_i}|^2+\varpi_i\rho_{R_i}|h_{R_iR_i}|^2+1},
  \end{equation}
where $\rho_{S_{i}}=\frac{P_S}{\sigma_{R_i}^2}$, $\rho_{R_i}=\frac{P_{R_i}}{\sigma_{R_i}^2}$, and $\sigma_{R_i}^2$ is the variance of the additive Gaussian noise at the relay $i$. Absolute value operator is represented by $|\cdot|$. SINR related to the signal of the second source at the $i^{th}$ relay is 
  \begin{equation} \label{Denklem_3}
  \gamma_{D_2}^{ R_i}=\frac{a_2 \rho_{S_{i}}|h_{S_i}|^2}{a_1\epsilon_{R_i}\rho_{S_{i}}|\tilde{g}_{S_i}|^2+\varpi\rho_{R_i}|h_{R_iR_i}|^2+1},
  \end{equation}
  where $\epsilon_{R_i}$ models the amount of the ipSIC at the $i^{th}$ relay and  $0\leq \epsilon_{R_i}\leq1$.
 
 The received signals at the destinations are
  \begin{equation} \label{Denklem_4}
  Y_{D_1}=\sqrt{a_3P_{R_i}}g_{R_{i1}}\hat{x}_1+\sqrt{a_4P_{R_i}}g_{R_{i1}}\hat{x}_2+n_{D_1}
  \end{equation}
and
  \begin{equation} \label{Denklem_5}
  Y_{D_2}=\sqrt{a_3P_{R_i}}h_{R_{i2}}\hat{x}_1+\sqrt{a_4P_{R_i}}h_{R_{i2}}\hat{x}_2+n_{D_2}.
  \end{equation}
 $n_{D_1}$ and  $n_{D_2}$ are the  additive Gaussian noise at the destinations.  The SINR related to the first source at the $D_1$ is
  \begin{equation} \label{Denklem_6}
  \gamma_{D_{1i}}^{ D_1}=\frac{a_3\rho_{D_{1i}}|g_{R_{i1}}|^2}{a_4\rho_{D_{1i}}|g_{R_{i1}}|^2+1},
  \end{equation}
 where $\rho_{D_{1i}}=\frac{P_{R_i}}{\sigma_{D_1}^2}$ and $\sigma_{D_1}^2$ is the noise variance at the $D_1$. Similarly, SINR related to the the first source at the $D_2$ is 
  \begin{equation} \label{Denklem_7}
  \gamma_{D_{1i}}^{ D_2}=\frac{a_3\rho_{D_{2i}}|h_{R_{i2}}|^2}{a_4\rho_{D_{2i}}|h_{R_{i2}}|^2+1},
  \end{equation}
and  SINR related to the second source at the $D_2$ is
  \begin{equation} \label{Denklem_8}
  \gamma_{D_{2i}}^{ D_2}=\frac{a_4\rho_{D_{2i}}|h_{R_{i2}}|^2}{a_3\epsilon_{D_2}\rho_{D_{2i}}|\tilde{h}_{R_{i2}}|^2+1}.
  \end{equation}
 $\rho_{D_{2i}}=\frac{P_{R_i}}{\sigma_{D_2}^2}$, $\sigma_{D_2}^2$ is the variance of the noise at the $D_2$, $\epsilon_{D_2}$ models the amount of the ipSIC at the  $D_2$ and  $0\leq \epsilon_{D_2}\leq1$.

\section{Relay Selection and Outage Probability Analysis} \label{OutageProbability} 

RS is carried out in two different methods, namely SSRS and TSRS, as offered in \cite{yue2018spatially}. 
 
SSRS is done by maximizing data rate of $D_1$ at the relay $i$ and two end-users: 
  \begin{equation} \label{Denklem_9}
  \begin{split}
  i_{S}=\arg \max_{\substack{i}} \{\min\{\log(1+\gamma_{D_1}^{R_i}), \log(1+\gamma_{D_{1i}}^{ D_1}), \log(1+\gamma_{D_{1i}}^{ D_2})\}, i \in\{1,\cdots, L\}\} 
  \end{split}.
  \end{equation}
   TSRS firstly determines a set of relays ensuring  service quality for $D_1$ and then selecting  a relay maximizing data rate of $D_2$. So the set of relays providing service quality priority condition is determined as follows:
  \begin{equation} \label{Denklem_10}
  \begin{split}
  K_R=\{\log(1+\gamma_{D_1}^{R_i})\geq R_{D_1}, \log(1+\gamma_{D_1i}^{ D_1}) \geq R_{D_1} , \log(1+\gamma_{D_1i}^{ D_2})\geq R_{D_1}\}, i \in\{1,\cdots, L\}\}. 
  \end{split}
  \end{equation}
  $R_{D_1}$ is the data rate of the $D_1$. Then, the relay maximizing data rate of $D_2$ is found as
 \begin{equation} \label{Denklem_11}
 \begin{split}
 i_{T}=\arg \max_{\substack{i}} \{\min\{\log(1+\gamma_{D_2}^{R_i}), \log(1+\gamma_{D_{2i}}^{ D_2})\}, i \in K_R\} 
 \end{split}.
 \end{equation}
Since a decision is made on selected relay in (\ref{Denklem_9}), (\ref{Denklem_10}), and (\ref{Denklem_11}), the prefactor $1/2$, which does not have any impact on the selected relay, is omitted for HD transmission.

\subsection{Outage Probability of Single Stage Relay Selection } \label{SSROP}

The OP of SSRS is (Please refer to Appendix A)
  \begin{equation} \label{Denklem_12}
  \begin{aligned}
    P_{SSRS}(\gamma_{th_1})&=Pr(\min\{\gamma_{D_1}^{R_{i_{S}}}, \gamma_{D_{1{i_{S}}}}^{D_1}, \gamma_{D_{1{i_{S}}}}^{D_2}\}<\gamma_{th_1})\\
    &=\prod_{i=1}^L \biggl(1-Pr(R_i \in K_R)\biggr),
      \end{aligned}
  \end{equation}
where $\gamma_{th_1}=2^{2R_{D_1}}-1$ for HD and   $\gamma_{th_1}=2^{R_{D_1}}-1$ for FD.
 $K_R$ represents the set of  relays providing service quality priority expressed in (\ref{Denklem_10}), a relay in that set satisfies $\min\{\gamma_{D_1}^{R_i}, \gamma_{D_{1i}}^{D_1}, \gamma_{D_{1i}}^{D_2}\}> \gamma_{th_1}$ which is expressed as  $R_i \in K_R$. Then,  $Pr(R_i \in K_R)$ is
 \begin{equation} \label{Denklem_13}
 \begin{split}
P_{R_i \in K_R}(\gamma_{th_1})=\frac{\Gamma\biggl(m_{D_{1i}}, \frac{m_{D_{1i}}\gamma_{th_1}}{\Omega_{D_{1i}}\rho_{D_{1i}}(a_3-\gamma_{th_1}a_4)} \biggr) \Gamma\biggl(m_{D_{2i}}, \frac{m_{D_{2i}}\gamma_{th_1}}{\Omega_{D_{2i}}\rho_{D_{2i}}(a_3-\gamma_{th_1}a_4)}\biggr) }{\Gamma(m_{D_{1i}})\Gamma(m_{D_{2i}})}  e^{-\frac{m_{S_{1i}}\gamma_{th_1}}{a_1 \rho_{S_{1i}}\Omega_{S_{1i}} }} \\  \sum_{p_1=0}^{m_{S_{1i}}-1} \sum_{p_2=0}^{p_1} \sum_{p_3=0}^{p_2} \biggl[ \biggl\{    \biggl(\frac{m_{S_{1i}}\gamma_{th_1}}{a_1\Omega_{S_{1i}}}\biggr )^{p_1}  \biggl(\frac{a_2\Omega_{S_{2i}}}{ m_{S_{2i}}}\biggr )^{p_3} \binom{p_1}{p_2}\binom{p_2}{p_3}  \Gamma(p_3+m_{S_{2i}})   \\ \Gamma(p_2-p_3+m_{R_i}) \biggl( \frac{\varpi_i\rho_{R_i}\Omega_{R_iR_i}}{m_{R_i}}  \biggr)^{p_2-p_3} \biggr\} \bigg/ \biggl\{ \Gamma(m_{S_{2i}}) \Gamma(m_{R_i})   \Gamma(p_1+1)     \\  \biggl(1+ \frac{m_{S_{1i}}\gamma_{th_1}a_2\Omega_{S_{2i}}}{m_{S_{2i}}a_1m_{S_{1i}} }  \biggr)^{m_{S_{2i}}+p_3}  \biggl(1+ \frac{m_{S_{1i}}\gamma_{th_1}\varpi_i \rho_{R_i} \Omega_{R_iR_i}}{m_{R_i}a_1  \rho_{S_i}\Omega_{S_{1i}} }  \biggr)^{m_{R_i}+p_2-p_3} (\rho_{S_i})^{p_1-p_3}      \biggr\} \biggr]  
\end{split}.
\end{equation}

At high SNRs, i.e, at high $\rho_{D_{1i}}$, $\rho_{D_{2i}}$, and $\rho_{S_i}$, asymptotic OP expression can be derived from (\ref{Denklem_13}) by considering the following facts:
\begin{itemize}
	\item [$\bullet$]  As $x \rightarrow0$, $e^{-x}\rightarrow 1$ and  $\gamma(m,x)  \rightarrow x^m$. Small $x$ values correspond to  the high SNRs cases.
	\item [$\bullet$]  At high SNRs,  $(1/\rho_{S_i})^{p_1-p_3}\rightarrow0$. Therefore, the only terms survive  in the summations of (\ref{Denklem_13}) are the ones where $p_1=p_3$, in turn, this equality means $p_1=p_2=p_3$. 
	\item [$\bullet$]  To observe effect of SI on OP at high SNRs, the term $\biggl(1+ \frac{m_{S_{1i}}\gamma_{th_1}\varpi_i \rho_{R_i} \Omega_{R_iR_i}}{m_{R_i}a_1  \rho_{S_i}\Omega_{S_{1i}} }  \biggr)^{m_{R_i}}$ is not canceled out, which approaches to zero since SI variance is assumed to be not only linearly but also exponentially dependent on the transmitted power of the relay. 
\end{itemize} 
Then, the asymptotic OP expression is obtained as
 \begin{equation} \label{Denklem_14}
 \begin{split}
P_{SSRS}^{\infty}(\gamma_{th_1})=\prod_{i=1}^L \biggl[1-\frac{\bigl( \frac{m_{D_{1i}}\gamma_{th_1}}{\Omega_{D_{1i}}\rho_{D_{1i}}(a_3-\gamma_{th_1}a_4)} \bigr)^{m_{D_{1i}}}}{\Gamma(m_{D_{1i}})} \biggr) \biggl( 1- \frac{\bigl( \frac{m_{D_{2i}}\gamma_{th_1}}{\Omega_{D_{2i}}\rho_{D_{2i}}(a_3-\gamma_{th_1}a_4)}\bigr)^{m_{D_{2i}}}}{\Gamma(m_{D_{2i}})}\biggr) \\   \sum_{p_1=0}^{m_{S_{1i}}-1}  \biggl[  \biggl\{  \binom{p_1+m_{S_{1i}}-1}{p_1}  \biggl(\frac{m_{S_{1i}}\gamma_{th_1}a_2\Omega_{S_{2i}}}{m_{S_{2i}}a_1\Omega_{S_{1i}}+m_{S_{1i}}\gamma_{th_1}a_2\Omega_{S_{2i}}}\biggr ) ^{p_1} \biggr\} \bigg/ \\ \biggl\{     \biggl(1+ \frac{m_{S_{1i}}\gamma_{th_1}a_2\Omega_{S_{2i}}}{m_{S_{2i}}a_1\Omega_{S_{1i}} }  \biggr)^{m_{S_{2i}}}    \biggl(1+ \frac{m_{S_{1i}}\gamma_{th_1}\varpi_i \rho_{R_i} \Omega_{R_iR_i}}{m_{R_i}a_1  \rho_{S_i}\Omega_{S_{1i}} }  \biggr)^{m_{R_i}}      \biggr\} \biggr]  \biggr]
\end{split}.
\end{equation}

The asymptotic expression in (\ref{Denklem_14}) reveals that at high SNRs, the parameters dominating OP are SI channel parameters (shape and scale (variance) parameters), PA factors at the sources, and fading channel parameters between the sources and the relays. On the other hand, PA factors at the relays and the channel parameters between the relays and destinations are only effective at the low SNRs.

\subsection{Outage Probability of Two Stage Relay Selection } \label{TSROP}

For TSRS, outage occurs in two cases, namely, when $K_R$ is an empty set or not, i.e., there is at least one relay satisfying condition given in (\ref{Denklem_10}) but mathematically no need to treat these states  separately. For any data rate of the second user, $R_{D_2}$, the OP of TSRS is calculated as
\begin{equation} \label{Denklem_15}
\begin{split}
 P_{TSRS}(\gamma_{th_1},\gamma_{th_2}) =
  &\sum_{l=0}^L Pr( \min(\gamma_{D_{2i_{T}}}^{D_2},\gamma_{D_2}^{R_{i_{T}}})< \gamma_{th_2}, |K_R|=l)\\
  =&\sum_{l=0}^L   \prod_{\substack{j=1 \\ R_i \in K_R}}^{l} \biggl( 1-P_{\phi}(\gamma_{th_1},\gamma_{th_2}) \biggr)
  \binom{L}{l} \prod_{\substack{j=1 \\ R_i \notin K_R}}^{L-l} \biggl(1- Pr(R_i\in K_R) \biggr)\\ 
  &\prod_{\substack{j=1 \\ R_i \in K_R}}^{l} \biggl( Pr(R_i\in K_R) \biggr)
\end{split}.
\end{equation}
The SINR threshold, $\gamma_{th_2}$, for HD and FD transmissions are $\gamma_{th_2}=2^{2R_{D_2}}-1$ and $\gamma_{th_2}=2^{R_{D_2}}-1$, respectively. The $|K_R|$ is the cardinality of $K_R$. $P_{\phi}(\gamma_{th_1},\gamma_{th_2})= Pr(\min(\gamma_{D_{2i}}^{D_2},\gamma_{D_2}^{R_i})> \gamma_{th_2} / \min\{\gamma_{D_1}^{R_i}, \gamma_{D_{1i}}^{D_1}, \gamma_{D_{1i}}^{D_2}\}> \gamma_{th_1})$ and it can be reformulated by considering the  Bayes theorem as
\begin{equation} \label{Denklem_16}
\begin{array} {ll}
P_{\phi}(\gamma_{th_1},\gamma_{th_2})=\frac{P_{\phi_1}(\gamma_{th_1},\gamma_{th_2})P_{\phi_2}(\gamma_{th_1},\gamma_{th_2})P_{\phi_3}(\gamma_{th_1},\gamma_{th_2})}{P_{R_i \in K_R}(\gamma_{th_1})}
\end{array},
\end{equation}
where $P_{\phi_1}(\gamma_{th_1},\gamma_{th_2})$, $P_{\phi_2}(\gamma_{th_1},\gamma_{th_2})$, and $P_{\phi_3}(\gamma_{th_1},\gamma_{th_2})$ are defined as in (\ref{B4}) and  their closed form expressions are given in (\ref{Denklem_17}), (\ref{Denklem_18}), and (\ref{Denklem_19}), respectively (Please refer to Appendix B).
 \begin{equation} \label{Denklem_17}
 \begin{split}
P_{\phi_1}(\gamma_{th_1},\gamma_{th_2})=1-\frac{\gamma\biggl(m_{D2i},  \frac{m_{D2i}U_{max}}{\Omega_{D_{2i}}}    \biggr)}{\Gamma(m_{D_{2i}})}-\frac{\biggl( \frac{m_{D_{2i}}\gamma_{th_2} a_3 \epsilon_{D_{2i}}  }{a_4 \tilde{m}_{D_{2i}}\Omega_{D_{2i}}} \biggr)^{m_{D_{2i}}}e^{\frac{\tilde{m}_{D_{2i}}}{a_3 \epsilon_{D_{2i}} \rho_{D_{2i}}}}}{\Gamma(m_{D_{2i}})   \biggl(1+ \frac{m_{D_{2i}}\gamma_{th_2} a_3 \epsilon_{D_{2i}}}{a_4 \tilde{m}_{D_{2i}}\Omega_{D_{2i}}}   \biggr)^{m_{D_{2i}}} } \\
\sum_{p_1=0}^{\tilde{m}_{D_{2i}}-1} \sum_{p_2=0}^{p_1} \biggl[ \frac{\binom{p_1}{p_2} \Gamma\biggl(p_2+m_{D_{2i}}, \bigl(\frac{m_{D_{2i}}}{\Omega_{D_{2i}}}+\frac{a_4 \tilde{m}_{D_{2i}}}{\gamma_{th_2}a_3\epsilon_{D_{2i}}} \bigr)U_{max} \biggr)    }{\Gamma(p_1+1) (-\rho_{D_{2i}})^{p_1-p_2} \bigl(  \frac{a_3\epsilon_{D_{2i}}}{\tilde{m}_{D_{2i}}} \bigr)^{p_1-p_2} \biggl(1+ \frac{m_{D_{2i}}\gamma_{th_2} a_3 \epsilon_{D_{2i}}}{a_4 \tilde{m}_{D_{2i}}\Omega_{D_{2i}}}   \biggr)^{p_2} }\biggr]
\end{split},
\end{equation}
where $U_{max}=\max\biggl(\frac{\gamma_{th_2}}{\rho_{D_{2i}}a_4}, \frac{\gamma_{th_1}}{\rho_{D_{2i}}(a_3-\gamma_{th_1}a_4)} \biggr)$.
 \begin{equation} \label{Denklem_18}
\begin{aligned}
P_{\phi_2}(\gamma_{th_1},\gamma_{th_2})=\left\{
\begin{array}{ll}
0 & \gamma_{th_1}\ge \frac{a_3}{a_4} \\
1-\frac{ \gamma\biggl(m_{D_{1i}},  \frac{m_{D_{1i}}\gamma_{th_1}}{\rho_{D_{1i}} \Omega_{D_{1i}}(a_3-\gamma_{th_1}a_4)}  \biggr)}{\Gamma(m_{D_{1i}}) } & \gamma_{th_1} < \frac{a_3}{a_4}\\
\end{array} 
\right.\\
\end{aligned}.
 \end{equation}
 \begin{equation} \label{Denklem_19}
 \begin{split}
P_{\phi_3}(\gamma_{th_1},\gamma_{th_2})=e^{- \frac{\gamma_{th_2} m_{S_{2i}a_1 \Omega_{S_{1i}}+ (\gamma_{th_2}+1) m_{S_{1i}}\gamma_{th_1} a_2 \Omega_{S_{2i}} }}{a_1  a_2 \rho_{S_i} \Omega_{S_{1i}} \Omega_{S_{2i}}}   }  \sum_{(q_1, q_2,q_3)=\overrightarrow{\rm \boldsymbol{0}} }^{(m_{S_{1i}}-1,q_1,q_2)} \sum_{(k_1, k_2,k_3)=\overrightarrow{\rm \boldsymbol{0}} }^{(q_3+m_{S_{2i}}-1,k_1,k_2)} \biggl[ \Phi_I \Phi_{II} \Phi_{III} \\ \frac{\bigl(\frac{\varpi_i \rho_{R_i}\Omega_{R_iR_i} }{m_{R_i}}\bigr)^{k_2+q_2-k_3-q_3}}{\bigl(\rho_{S_i}\bigr)^{k_1+q_1-k_3-q_3}} \biggr]
\end{split},
\end{equation}
where $ \sum_{(q_1, q_2,q_3)=\overrightarrow{\rm \boldsymbol{0}} }^{(m_{S_{1i}}-1,q_1,q_2)} \sum_{(k_1, k_2,k_3)=\overrightarrow{\rm \boldsymbol{0}} }^{(q_3+m_{S_{2i}}-1,k_1,k_2)}= \sum_{q_1=0 }^{m_{S_{1i}}-1}   \sum_{q_2=0 }^{q_1} \sum_{q_3=0 }^{q_2} \sum_{k_1=0}^{q_3+m_{S_{2i}}-1}\sum_{k_2=0}^{k_1}\sum_{k_3=0}^{k_2}$. $\overrightarrow{\rm \boldsymbol{0}}$ is a three dimensional zero vector and  $ \Phi_I$, $\Phi_{II}$, and $\Phi_{III}$ in expression of $P_{\phi_3}(\gamma_{th_1},\gamma_{th_2})$ are
\begin{equation} \label{Denklem_20}
\begin{split}
\Phi_I=&\frac{\binom{k_1}{k_2} \binom{k_2}{k_3} \binom{q_1}{q_2} \binom{q_2}{q_3} \Gamma(k_3+\tilde{m}_{R_i}) \Gamma(k_2+q_2+m_{R_i}-k_3-q_3) \Gamma(q_3+m_{S_{2i}}) \bigl(\frac{m_{S_{1i}}\gamma_{th_1} }{a_1\Omega_{S_{1i}}} \bigr)^{q_1}    }{\Gamma(m_{S_{2i}}) \Gamma(m_{R_i}) \Gamma(\tilde{m}_{R_i}) \Gamma(k_1+1) \Gamma(q_1+1)} 
\\&\biggl(\frac{a_2\Omega_{S_{2i}}}{m_{S_{2i}}} \biggr)^{q_3-k_1}\biggl(\frac{a_1\epsilon_{R_i}}{\tilde{m}_{R_i}} \biggr)^{k_3} (\gamma_{th_2})^{k_1}
\\
\Phi_{II}=& \frac{1}{\biggl(1+ \frac{m_{S_{1i}} \gamma_{th_1} a_2 \Omega_{S_{2i}}  }{m_{S_{2i}}a_1 \Omega_{S_{1i}} }\biggr)^{q_3+m_{S_{2i}}-k_1} \biggl(1+ \frac{\gamma_{th_2} (m_{S_{2i}} a_1 \Omega_{S_{1i}} + m_{S_{1i}} \gamma_{th_1} a_2 \Omega_{S_{2i}}   )}{\tilde{m}_{R_i} a_2 \Omega_{S_{1i}} \Omega_{S_{2i}}} \epsilon_{R_i}   \biggr)^{k_3+\tilde{m}_{R_i}}  }
\\
\Phi_{III}=& \frac{1}{ \biggl( 1 + \biggl( \frac{m_{S_{2i}} a_1 \Omega_{S_{1i}} \gamma_{th_2} +  (\gamma_{th_2}+1) m_{S_{1i}}\gamma_{th_1} a_2 \Omega_{S_{2i}}   }{m_{R_i}a_1 a_2 \Omega_{S_{1i}} \Omega_{S_{2i}}}     \biggr)  \frac{\varpi_i \rho_{R_i} \Omega_{R_iR_i} }{\rho_{S_i}}  \biggr)^{k_2+q_2+m_{R_i}-k_3-q_3}   }
\end{split}.
\end{equation}
The closed form of $Pr(R_i\in K_R)$ is obtained in the derivation of $P_{SSRS}(\gamma_{th_1})$, as given in (\ref{Denklem_13}).

Using the same approach as in derivation of $P_{SSRS}^{\infty}(\gamma_{th_1})$,  asymptotic  expressions for $P_{\phi_1}(\gamma_{th_1},\gamma_{th_2})$, $P_{\phi_2}(\gamma_{th_1},\gamma_{th_2})$, $P_{\phi_3}(\gamma_{th_1},\gamma_{th_2})$,  and $P_{R_i \in K_R}^{\infty}(\gamma_{th_1})$ are derived as follows
 \begin{equation} \label{Denklem_21}
 \begin{split}
P_{\phi_1}^{\infty}(\gamma_{th_1},\gamma_{th_2})=1-\sum_{p_1=0}^{\tilde{m}_{D_{2i}}-1} \biggl[ \frac{\Gamma\bigl(p_1+m_{D_{2i}} \bigr) \biggl( \frac{m_{D_{2i}}\gamma_{th_2} a_3 \epsilon_{D_{2i}}  }{a_4 \tilde{m}_{D_{2i}}\Omega_{D_{2i}}} \biggr)^{m_{D_{2i}}}   }{\Gamma(p_1+1)  \Gamma(m_{D_{2i}})   \biggl(1+ \frac{m_{D_{2i}}\gamma_{th_2} a_3 \epsilon_{D_{2i}}}{a_4 \tilde{m}_{D_{2i}}\Omega_{D_{2i}}}   \biggr)^{m_{D_{2i}}+p_2} }\biggr]
\end{split},
\end{equation}

 \begin{equation} \label{Denklem_22}
 P_{\phi_2}^{\infty}(\gamma_{th_1},\gamma_{th_2})=1- \frac{1}{\Gamma(m_{D_{1i}}+1)} \biggl(\frac{m_{D_{1i}}\gamma_{th_1}}{\rho_{D_{1i}} \Omega_{D_{1i}}(a_3-\gamma_{th_1}a_4)} \biggr)^{m_{D_{1i}}},
 \end{equation}

 \begin{equation} \label{Denklem_24}
 \begin{split}
P_{\phi_3}^{\infty}(\gamma_{th_1},\gamma_{th_2})=\sum_{q_1=0}^{m_{S_{1i}}-1} \sum_{k_1=0}^{q_1+m_{S_{1i}}-1} \biggl[ \biggl\{ \Gamma(\tilde{m}_{R_i}+k_1) \Gamma(m_{S_{2i}}+q_1)  \biggl( \frac{m_{S_{2i}}}{a_2 \Omega_{S_{2i}}}   \biggr)^{k_1-q_1}  \biggl( \frac{m_{S_{1i}} \gamma_{th_1} }{a_1 \Omega_{S_{1i}}}   \biggr)^{q_1}  \\  \biggl( \frac{ \gamma_{th_2}a_1 \epsilon_{R_i} }{ \tilde{m}_{R_i}}   \biggr)^{k_1}   \biggr \}     \bigg/ \biggl\{ \Gamma(m_{S_{2i}}) \Gamma(\tilde{m}_{R_i}) \Gamma(k_1+1) \Gamma(q_1+1) \biggl(1+ \frac{m_{S_{1i}} \gamma_{th_1} a_2 \Omega_{S_{2i}}  }{m_{S_{2i}}a_1 \Omega_{S_{1i}} }\biggr)^{m_{S_{2i}}+q_1-k_1} \\  \biggl( 1 + \biggl( \frac{m_{S_{2i}} a_1 \Omega_{S_{1i}} \gamma_{th_2} +  (\gamma_{th_2}+1) m_{S_{1i}}\gamma_{th_1} a_2 \Omega_{S_{2i}}   }{m_{R_i}a_1 a_2 \Omega_{S_{1i}} \Omega_{S_{2i}}}     \biggr)  \frac{\varpi_i \rho_{R_i} \Omega_{R_iR_i} }{\rho_{S_i}}  \biggr)^{m_{R_i}} \\ \biggl(1+ \frac{\gamma_{th_2} (m_{S_{2i}} a_1 \Omega_{S_{1i}} + m_{S_{1i}} \gamma_{th_1} a_2 \Omega_{S_{2i}}   )}{\tilde{m}_{R_i} a_2 \Omega_{S_{1i}} \Omega_{S_{2i}}} \epsilon_{R_i}   \biggr)^{\tilde{m}_{R_i}+k_1}     \biggr \}     \biggr]
\end{split},
\end{equation}
and
   \begin{equation} \label{Denklem_23}
   \begin{split}
  P_{R_i \in K_R}^{\infty}(\gamma_{th_1})=\biggl( 1- \frac{\biggl( \frac{m_{D_{1i}}\gamma_{th_1}}{\Omega_{D_{1i}}\rho_{D_{1i}}(a_3-\gamma_{th_1}a_4)} \biggr)^{m_{D_{1i}}}}{\Gamma(m_{D_{1i}})} \biggr) \biggl( 1- \frac{\biggl( \frac{m_{D_{2i}}\gamma_{th_1}}{\Omega_{D_{2i}}\rho_{D_{2i}}(a_3-\gamma_{th_1}a_4)}\biggr)^{m_{D_{2i}}}}{\Gamma(m_{D_{2i}})}\biggr) \\   \sum_{p_1=0}^{m_{S_{1i}}-1}  \biggl[  \biggl\{  \binom{p_1+m_{S_{1i}}-1}{p_1}  \biggl(\frac{m_{S_{1i}}\gamma_{th_1}a_2\Omega_{S_{2i}}}{m_{S_{2i}}a_1\Omega_{S_{1i}}+m_{S_{1i}}\gamma_{th_1}a_2\Omega_{S_{2i}}}\biggr ) ^{p_1} \biggr\} \bigg/ \\ \biggl\{     \biggl(1+ \frac{m_{S_{1i}}\gamma_{th_1}a_2\Omega_{S_{2i}}}{m_{S_{2i}}a_1\Omega_{S_{1i}} }  \biggr)^{m_{S_{2i}}}    \biggl(1+ \frac{m_{S_{1i}}\gamma_{th_1}\varpi_i \rho_{R_i} \Omega_{R_iR_i}}{m_{R_i}a_1  \rho_{S_i}\Omega_{S_{1i}} }  \biggr)^{m_{R_i}}      \biggr\} \biggr] 
  \end{split}.
  \end{equation}
Consequently, the asymptotic expression of   $P_{TSRS}(\gamma_{th_1},\gamma_{th_2})$ is
\begin{equation} \label{Denklem_25}
\begin{split}
P_{TSRS}^{\infty}(\gamma_{th_1},\gamma_{th_2})=&\sum_{l=0}^L   \prod_{\substack{j=1 \\ R_i \in K_R}}^{l} \biggl( 1-P_{\phi}^{\infty}(\gamma_{th_1},\gamma_{th_2}) \biggr)
\binom{L}{l} \prod_{\substack{j=1 \\ R_i \notin K_R}}^{L-l} \biggl(1- P_{R_i \in K_R}^{\infty}(\gamma_{th_1})\biggr)\\ 
&\prod_{\substack{j=1 \\ R_i \in K_R}}^{l} \biggl(P_{R_i \in K_R}^{\infty}(\gamma_{th_1}) \biggr)
\end{split},
\end{equation}
where $P_{\phi}^{\infty}(\gamma_{th_1},\gamma_{th_2})=P_{\phi_1}^{\infty}(\gamma_{th_1},\gamma_{th_2})P_{\phi_2}^{\infty}(\gamma_{th_1},\gamma_{th_2})P_{\phi_3}^{\infty}(\gamma_{th_1},\gamma_{th_2})/P_{R_i \in K_R}^{\infty}(\gamma_{th_1})$.

For pSIC and HD schemes, numerical results can be obtained from the above equations. Although $1/0$ takes place in (\ref{Denklem_17}) can be eliminated by using limit functions in software programs such as MATHEMATICA and MAPLE, we provide  its pSIC counterpart as in (\ref{Denklem_26}). Simplification of other equations for pSIC and HD cases are straightforward and actually they can be reached by  cancellation of the terms including  $\epsilon_{R_i}$, $\epsilon_{D_{2i}}$, and $\varpi_i$. 

\begin{equation} \label{Denklem_26}
\begin{aligned}
P_{\phi_1}^{pSIC}(\gamma_{th_1},\gamma_{th_2})&=Pr\biggl( |h_{R_{i2}}|^2 > \frac{\gamma_{th_1}}{a_3\rho_{D_{2i}}-a_4\rho_{D_{2i}}\gamma_{th_1}}  \biggr)\\
&=1-\frac{\gamma\biggl(m_{D2i},  \frac{m_{D2i}U_{max}}{\Omega_{D_{2i}}}    \biggr)}{\Gamma(m_{D_{2i}})}
\end{aligned}
\end{equation}

\section{Numerical Results} \label{NumericalResults}

In this section, numerical results for OP are provided to validate the derived expressions. It is assumed that  $P_T=2P_S=2P_{R_i}$ and $\sigma^2=\sigma_{R_i}^2=\sigma_{D_1}^2=\sigma_{D_2}^2=1$. The path loss exponent, representing urban environment, is taken to be 3. The curves are plotted versus $P_T/\sigma^2$.

\begin{figure}[htbp] 
 	\centering
 	\includegraphics[width=4.4in]{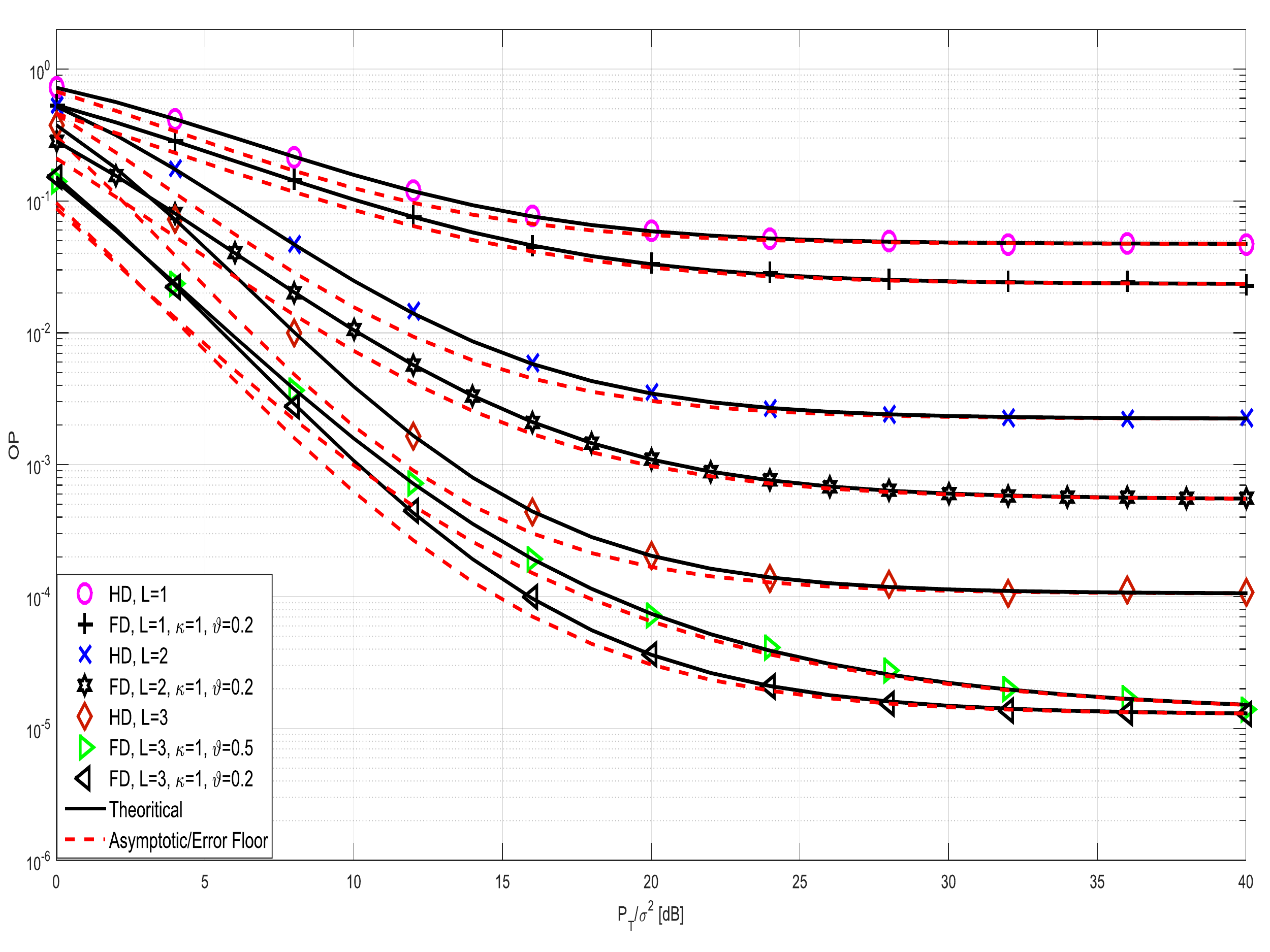}
 	\caption{Exact,  asymptotic, and simulation OPs of SSRS over Rayleigh fading channels.}
 	\label{OPofSSRSOverRayliehgFade}
 \end{figure}

 Fig. \ref{OPofSSRSOverRayliehgFade} verifies correctness of exact and asymptotic expressions of OP for SSRS strategy over Rayleigh fading environment, namely, (\ref{Denklem_12}) and (\ref{Denklem_14}). PAs are chosen as   $a_1=a_3=0.75$ ve $a_2=a_4=0.25$.  In meanwhile variances of channel gains are set to $1$,  and data rates are $R_{D_1}=0.1$ bit per channel use (BPCU)  and $R_{D_2}=1$ BPCU.  SI cancellation factors are chosen as $(\kappa_l, \vartheta_l) \in \{(1, 0.2), (1, 0.5)\}$. Simulation and analytical results coincide in excellent way for both FD and HD transmissions and FD transmission provides better performance in case of sufficient SI cancellation. Effect of SI is also illustrated, however, as SNR ($P_T/\sigma^2$) increases the same error floor is reached due to SIC effect which dominates performance even if there are different SI powers. Furthermore, as number of relays increases OP decreases, i.e., substantial gain is attained. As $P_T/\sigma^2$ goes to infinity SINRs approach constant values, therefore, error floors occur. But error floor reduces as total number of relays increases, i.e., diversity order demonstrates its effect on the OP. In error floor region diversity order is not zero but due to the constant SINR values even if $P_T/\sigma^2$ increases, they attain fixed value and this results in an error floor. Unlike HD systems where OP or SERs are altmost proportional to the transmitted power, in turn SNR, in the systems where FD or NOMA transmission is implemented, error probabilities are commonly not proportional to the SNR. For example, in FD transmission, error floor  takes place if SI power is only linearly dependent on the transmitted power, which causes a fixed SINR value as $P_T/\sigma^2$ reaches infinity. However, no error floor is observed if SI power  also exponentially depends on transmitted power as illustrated in this work. Actually, linearly or exponentially dependence of SI power is not the reason for zero or non-zero diversity order. Misinterpretation of diversity order as slope of log-log plot causes this undesired situation although diversity order is the number of independent paths from the source to the destination. 
  
\begin{figure}[htbp] 
	\centering
	\includegraphics[width=4.4in]{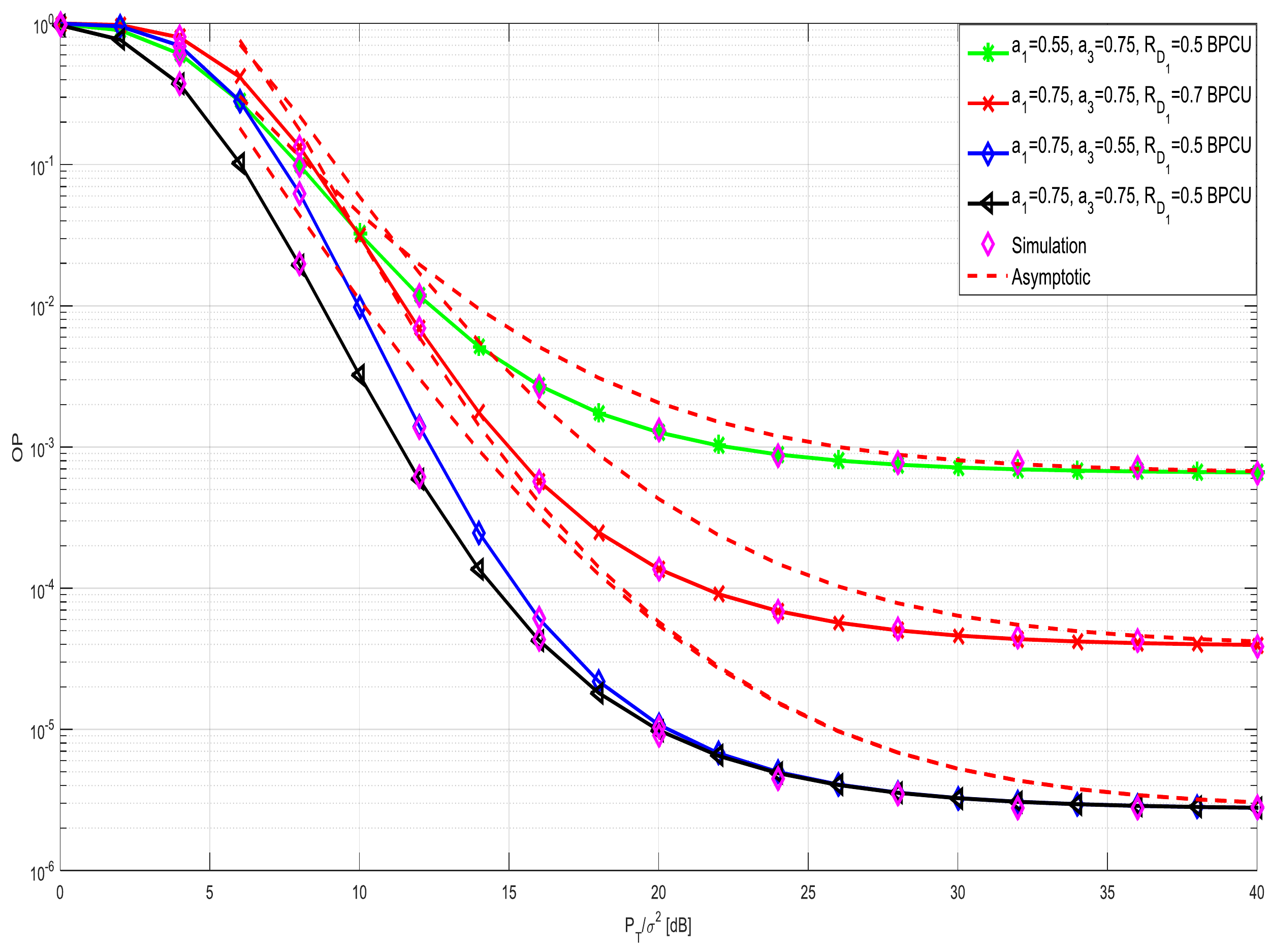}
	\caption{Exact,  asymptotic, and simulation OPs of SSRS over Nakagami-$m$ fading channels.}
	\label{OPofSSRSOverNakagamimFade}
\end{figure}

 In Fig. \ref{OPofSSRSOverNakagamimFade}, validity of exact and asymptotic expressions given in  (\ref{Denklem_12}) and (\ref{Denklem_14}) for Nakagami-$m$ fading channels for FD transmission is investigated, where all shaping parameters are set to 2 and scaling parameters (variances) are set to 1, thereby,  distances are also equal to 1. Results are demonstrated for different PAs ($a_1 \in \{0.55, 0.75\}$ \& $a_3 \in \{0.55, 0.75\}$)  and data rates of user 1, $R_{D_1} \in \{0.5, 0.7\}$. The data rate of user 2 is set to $R_{D_2}=1$, total number of relays is $L=4$, and SI cancellation factors are chosen as $(\kappa_l, \vartheta_l)=(1, 0.2)$. The overlap of the simulation and exact results is excellent and asymptotic ones follow them. Error floor of two lower curves is the same even different powers are assigned to the user 1 at the relays. This result is not surprising since power allocation at the sources but not at the relays is effective on error floor, which is obvious as can be observed from (\ref{Denklem_14}).

\begin{figure}[htbp] 
	\centering
	\includegraphics[width=4.4in]{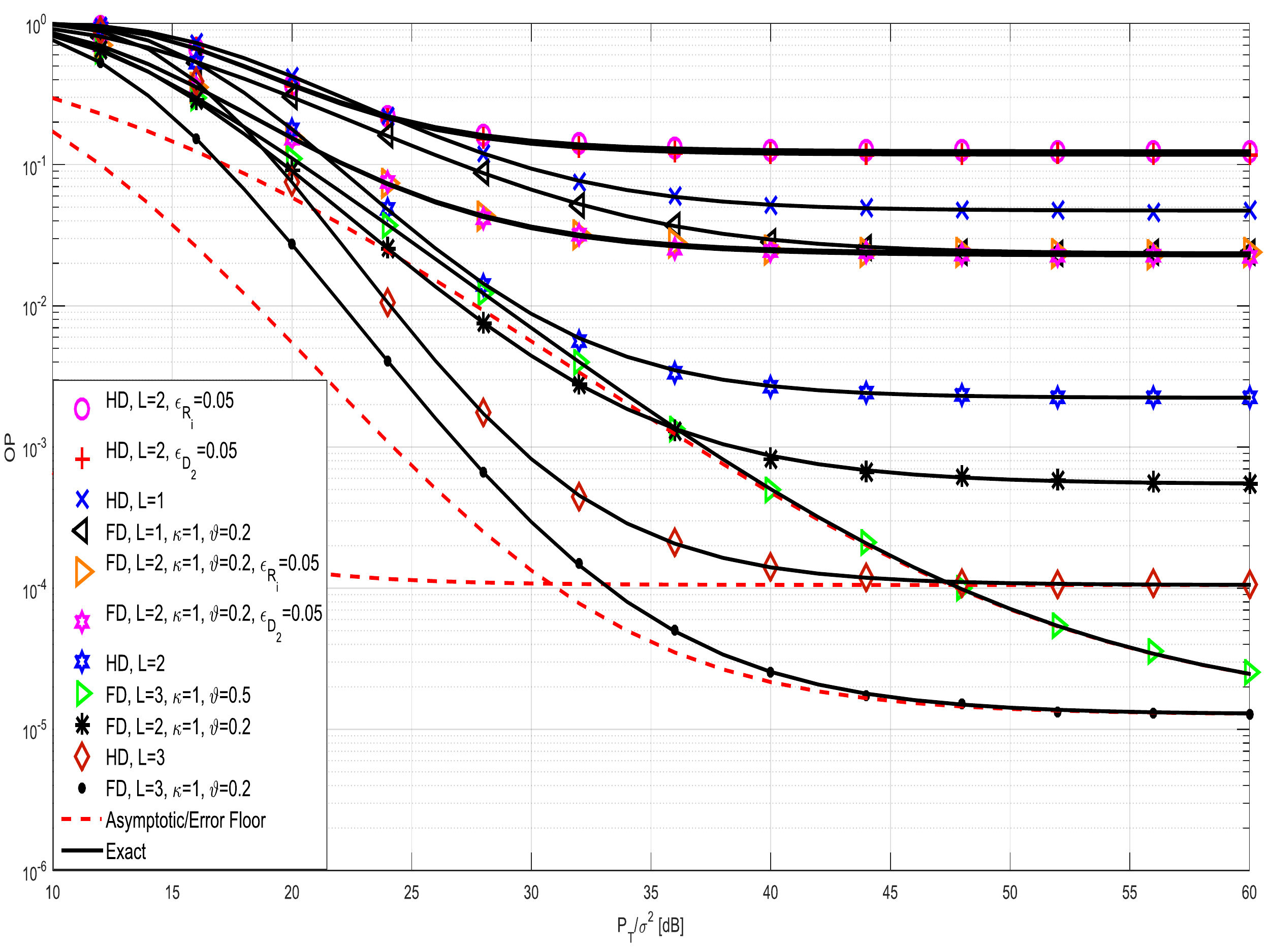}
	\caption{Exact, asymptotic, and simulation OPs of TSRS over Rayleigh fading channels.}
	\label{OPofTSRSOverRayliehgFade}
\end{figure}

Fig. \ref{OPofTSRSOverRayliehgFade} demonstrates results of OP for TSRS over Rayleigh fading environment for both HD and FD transmissions, where all parameter settings are kept the same as in Fig. \ref{OPofSSRSOverRayliehgFade}. Additionally, to illustrate effect of ipSIC on performance, two schemes for both HD and FD protocols are considered where $\epsilon_{R_i}=\epsilon_{D_{2i}}=0.05$ for $L=2$. For the sake of figure clarity, asymptotic curves are only provided for $L=3$. Exact and asymptotic curves are derived from (\ref{Denklem_15}) and (\ref{Denklem_25}), respectively. Exact and simulation results perfectly coincide with each other. Asymptotic curves excellently overlap with exact and simulation curves in error floors. All deductions in Fig. \ref{OPofSSRSOverRayliehgFade} for SSRS are also valid for TSRS. Unlike SSRS, SI is more effective where worse performance can be attained in case of insufficient SI cancellation.  Imperfect SIC at the relay and destination has nearly the same effect and it reduces performance significantly.  

\begin{figure}[htbp] 
	\centering
	\includegraphics[width=4.4in]{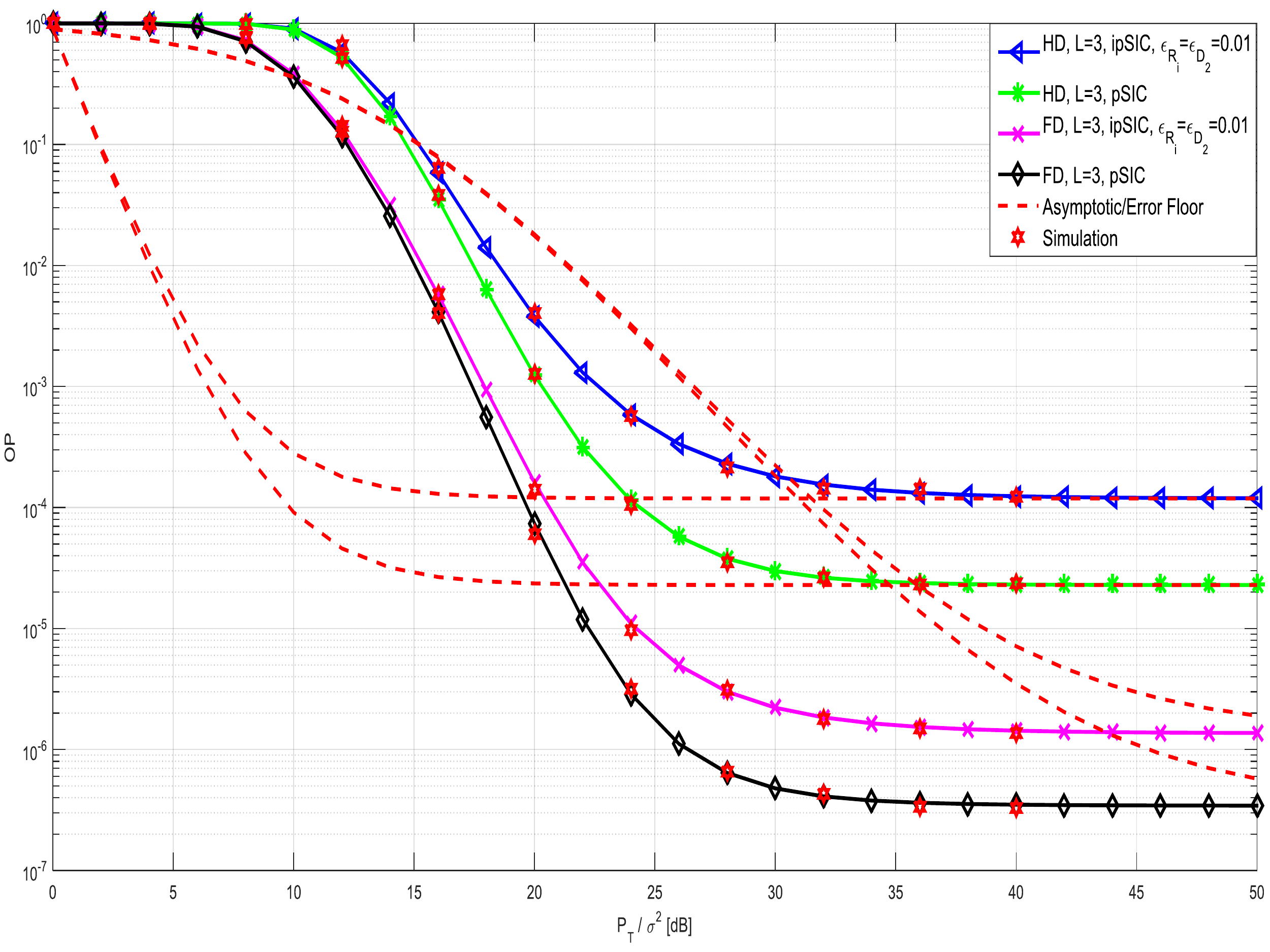}
	\caption{Exact, asymptotic, and simulation OPs of TSRS over Nakagami-$m$ fading channels.}
	\label{OPofTSRSOverNakagami_m}
\end{figure} 

Fig. \ref{OPofTSRSOverNakagami_m} verifies correctness of the OP expression of TSRS for different shape/scale factors and also for imperfect SIC cases.   Error floor happens due to SIC, imperfect SIC, and insufficient SI cancellation. The shape and  scale parameters are chosen as $(m_{S_{1i}}=2, \Omega_{S_{1i}}=1.55)$, $(m_{S_{2i}}=3,\Omega_{S_{2i}}=1.7)$, $(m_{D_{1i}}=1.75, \Omega_{D_{1i}}=1.3)$, $(m_{D_{2i}}=1.5, \Omega_{D_{2i}}=1.6)$, $(\tilde{m}_{R_{i}}=2, 1)$, $(\tilde{m}_{D_{2i}}=2, 1)$, and $(m_{R_iR_i}=1.25, \Omega_{R_iR_i}=P_{R_i}^{0.2})$. Data rates are $R_{D_1}=0.2$ BPCU and $R_{D_2}=1$ BPCU. Total number of relays is $L=3$ and ipSIC parameters are $\epsilon_{D_{2i}}=\epsilon_{R_i}=0.01$. PA parameters are $a_1=a_3=0.75$. Overlapping of exact and simulation results reveals validity of OP expressions for non-integer shaping parameters, namely, $m_{D_{1i}}$, $m_{D_{2i}}$, $\tilde{m}_{R_{i}}$, and $m_{R_iR_i}$, whereas other three ones are assumed to be integers. Although ipSIC parameters are set to a small value, $1 \%$, their effect on performance is enormous. The OP becomes about  $9$ times of the former value for both  HD and FD transmission, at a $P_T / \sigma^2$ value of 40 dB.

\begin{figure}[htbp] 
	\centering
	\includegraphics[width=4.4in]{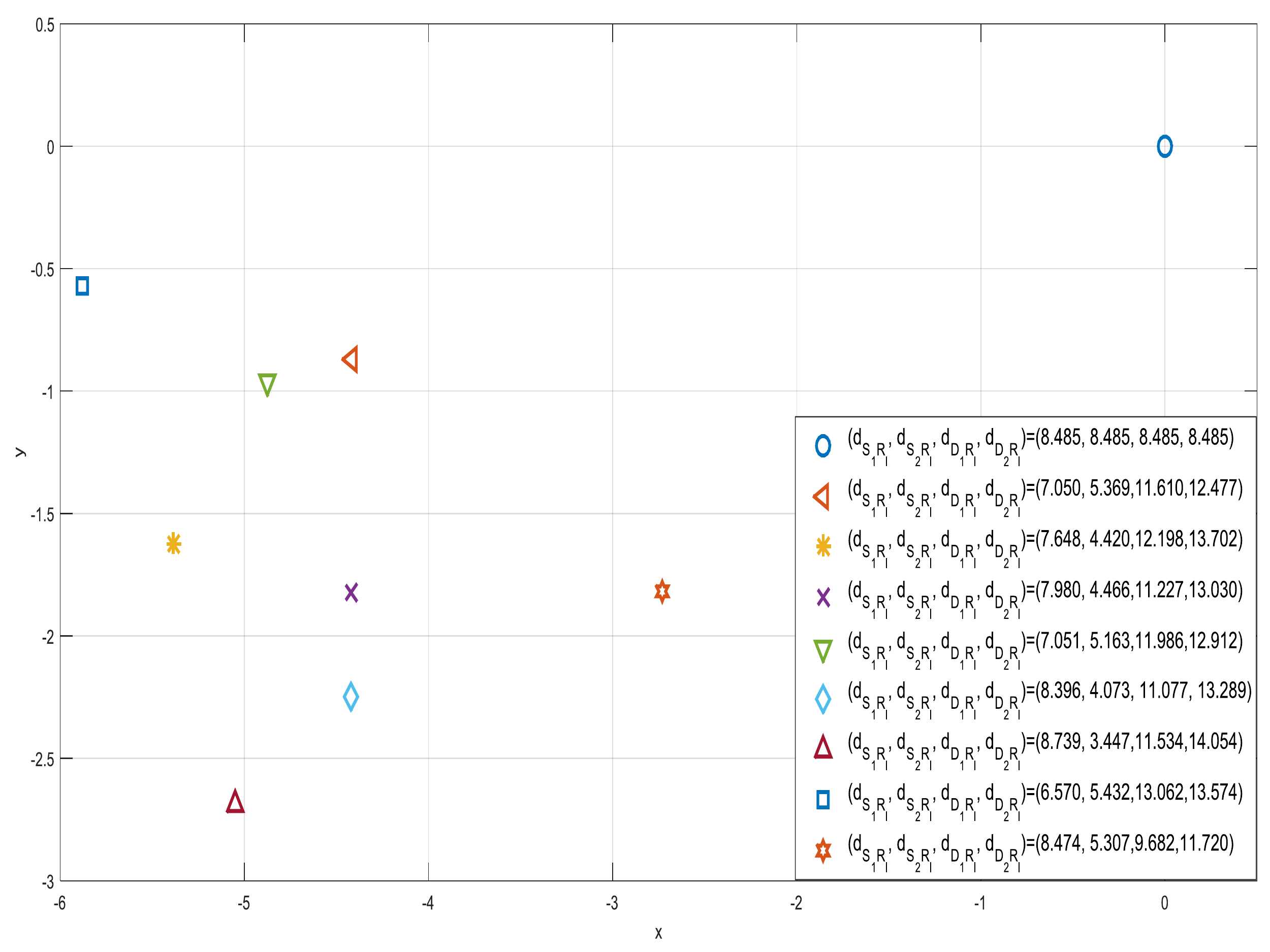}
	\caption{Optimum relay locations of FD transmission.}
	\label{FDOPRelayLocationXY}
\end{figure}

\begin{figure}[htbp] 
	\centering
	\includegraphics[width=4.4in]{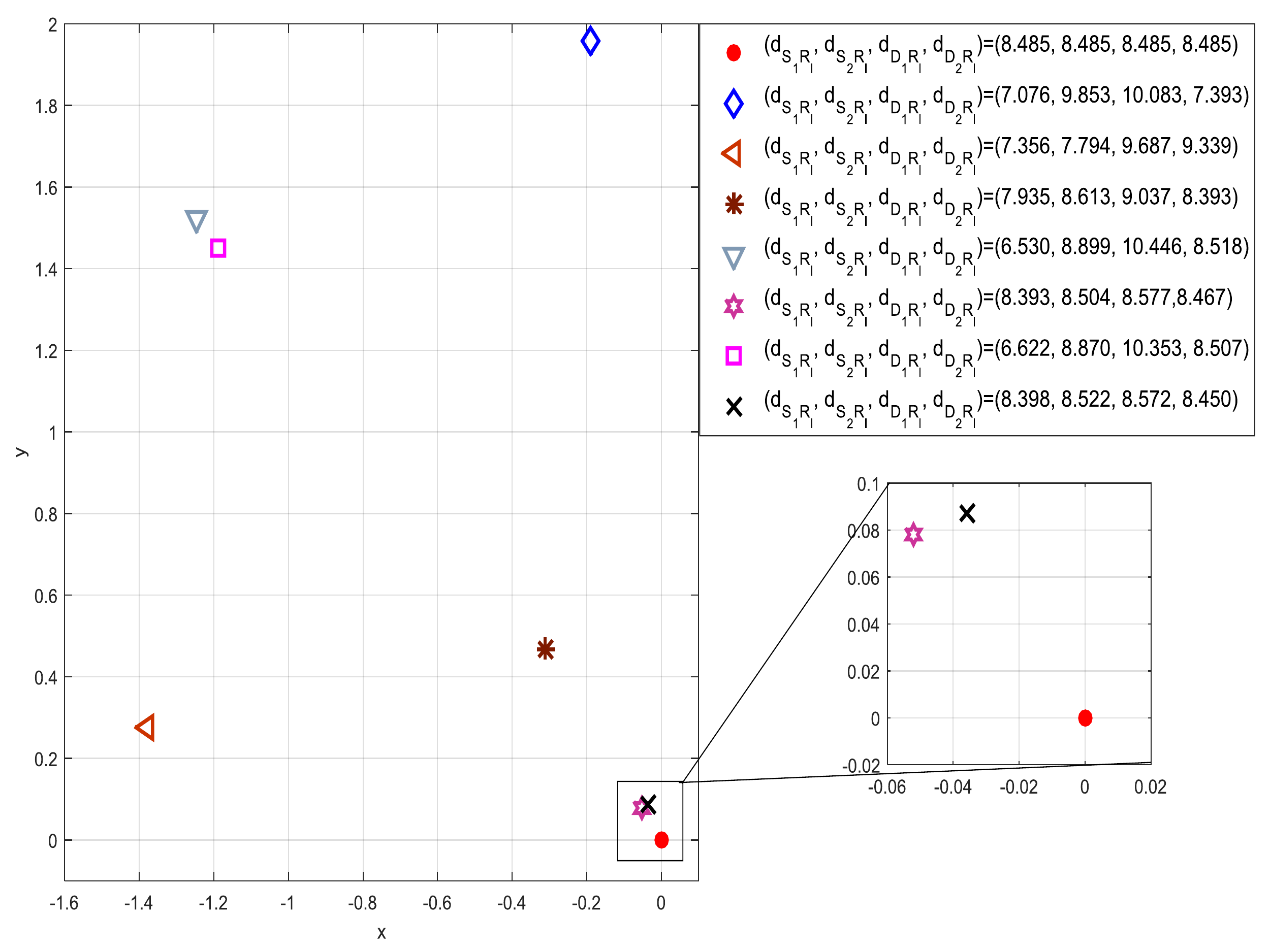}
	\caption{Optimum relay locations of HD transmission.}
	\label{HDOPRelayLocationXY}
\end{figure}

Tables \ref{RelayLocationResultsFD}, \ref{FDOptimumDistance}, \ref{RelayLocationResultsHD}, \ref{HDOptimumDistance} together with Fig. \ref{FDOPRelayLocationXY} and \ref{HDOPRelayLocationXY} illustrate the performance of different schemes of TSRS over Nakagami-$m$ fading channels for optimum relay locations. All shape factors are set to 2 and scaling factors (variances) are assumed to be only characterized by relay location and path loss exponent, $\alpha$. The nodes $S_1$, $S_2$, $D_1$, and $D_2$ are assumed to be located at the points $(x_{S_1}, y_{S_1})=(-6, 6)$, $(x_{S_2}, y_{S_2})=(-6, -6)$, $(x_{D_1}, y_{D_1})=(6, -6)$, and $(x_{D_2}, y_{D_2})=(6, 6)$, respectively, which corresponds to the case of equal distances among the relays and nodes. As assumed, the relays are clustered and located at the origin, thereafter, different schemes are assigned and related optimum relay locations are obtained. The relay's location is set to $\bigl(x_{R_l}, y_{R_l}\bigr)=\bigl(r\cos(\theta), r\sin(\theta)\bigr)$ where  $r=n\times6/64$, $n:=0:1:63$ and $\theta=p\times6.28/64$, $p:=0:1:63$. All coordinates are in meters and first three rows are not optimum relay locations but provided for comparison. Table \ref{FDOptimumDistance} includes distances among each node and relays for the schemes provided in Table \ref{RelayLocationResultsFD} for FD transmission. Similarly, Table \ref{HDOptimumDistance} includes distances corresponding to the  schemes provided in Table \ref{RelayLocationResultsHD} for HD transmission. For FD transmission, all optimum locations fall in the third quarter of the cartesian system, i.e. relays  move towards the second source, on the other hand, those of HD transmission are located in the second quarter and their changes with respect to origin are not as hard as that of FD.  This interesting finding is due to SI effect which governs the best relay location of the FD transmission. Unequal power allocations at the sources and the relays result in  better performance than equal allocations for both FD and HD protocols. Increasing total transmitted power in FD transmission moves the best relay location towards the middle of the sources. The best relay location of FD transmission is more sensitive to PA factors, users' data rates, ipSICs at the relay and user 2, and total power, whereas the data rate of the second user and ipSICs are not more effective in HD transmission.

\begin{table}[]
	\centering
	\caption{Optimum relay location of FD transmission for TSRS.}
	\begin{tabular}{|c|c|c|c|c|c|c|c|}
		\hline
		\textbf{\begin{tabular}[c]{@{}c@{}}  $\boldsymbol{P_T/\sigma^2}$\\ (dB)\end{tabular}} & $\boldsymbol{a_1}$ & $\boldsymbol{a_3}$ & $(\boldsymbol{\kappa_l}, \boldsymbol{\vartheta_l})$ & $(\boldsymbol{\epsilon_{R_l}}, \boldsymbol{\epsilon_{D_{2l}}})$ & $(\boldsymbol{R_{D_l}}, \boldsymbol{R_{D_2}})$  & $(\boldsymbol{x_{R_l}}, \boldsymbol{y_{R_l}})$ & $\boldsymbol{OP_{\min}}$ \\ \hline
		50 & 0.55 & 0.75  & (1, 0.31)  & (0, 0) & (0.1, 1) & (0.000, 000) & $3.69\times 10^{-2}$ \\ \hline
		50 & 0.75  & 0.55 & (1, 0.31) & (0, 0) & (0.1, 1) & (0.000, 000) & $1.74\times 10^{-1}$ \\ \hline
		50 & 0.75 & 0.75   & (1, 0.31) & (0, 0) & (0.1, 1)  & (0.000, 000) & $1.75\times 10^{-1}$ \\ \hline
		50 & 0.55  & 0.75 & (1, 0.31) & (0, 0) & (0.1, 1) & (-4.415, -0.870) & $6.27\times 10^{-4}$ \\ \hline
		50 & 0.75 & 0.55 & (1, 0.31) & (0, 0) & (0.1, 1) & (-5.386, -1.623) & $3.70\times 10^{-4}$  \\ \hline
		50  & 0.75 & 0.75 & (1, 0.31) & (0, 0) & (0.1, 1) & (-4.421, -1.822) & $9.10\times 10^{-4}$ \\ \hline
		50 & 0.75 & 0.75 & (1, 0.31) & (0, 0)   & (0.3, 1) & (-4.875, -0.961) & $6.58\times 10^{-3}$ \\ \hline
		50 & 0.75 & 0.75 & (1, 0.31) & (0, 0) & (0.1, 2) & (-4.421, -2.246) & $8.52\times 10^{-2}$ \\ \hline
		50  & 0.75 & 0.75 & (1, 0.51) & (0, 0) & (0.1, 1)  & (-5.048, -2.687)  & $2.19\times 10^{-2}$  \\ \hline
		60 & 0.75 & 0.75 & (1, 0.31) & (0, 0) & (0.1, 1) & (-5.879, -0.569) & $1.70\times 10^{-7}$  \\ \hline
		50 & 0.75 & 0.75 & (1, 0.31) & $(10^{-4}, 10^{-4})$ & (0.1, 1) & (-2.732, -1.818) & $5.38\times 10^{-2}$ \\ \hline
	\end{tabular}
\label{RelayLocationResultsFD}
\end{table}

\begin{table}[]
	\centering
	\caption{Optimum distances of FD transmission for TSRS.}
	\begin{tabular}{|c|c|c|c|c|}
		\hline
		$\boldsymbol{d_{S_1R_l}}$ & $\boldsymbol{d_{S_2R_l}}$ & $\boldsymbol{d_{D_1R_l}}$ & $\boldsymbol{d_{D_2R_l}}$ & $\boldsymbol{OP_{\min}}$\\ \hline
		8.485 &	8.485 &	8.485 &	8.485  & $3.69\times 10^{-2}$  \\ \hline
		8.485 &	8.485 &	8.485 &	8.485  & $1.74\times 10^{-1}$ \\ \hline
		8.485 &	8.485 &	8.485 &	8.485  & $1.75\times 10^{-1}$ \\ \hline
		7.050 &	5.369 &	11.610 & 12.477 & $6.27\times 10^{-4}$  \\ \hline
		7.648 &	4.420 &	12.198 & 13.702 & $3.70\times 10^{-4}$  \\ \hline
		7.980 &	4.466 &	11.227 & 13.030 & $9.10\times 10^{-4}$  \\ \hline
		7.051 &	5.163 &	11.986 & 12.912 & $6.58\times 10^{-3}$ \\ \hline
		8.396 &	4.073 &	11.077 & 13.289  & $8.52\times 10^{-2}$  \\ \hline
		8.739 &	3.447 &	11.534 & 14.054 & $2.19\times 10^{-2}$   \\ \hline
		6.570 &	5.432 &	13.062 & 13.574 & $1.70\times 10^{-7}$   \\ \hline
		8.474 &	5.307 &	9.682 &	11.720  & $5.38\times 10^{-2}$  \\ \hline
	\end{tabular}
	\label{FDOptimumDistance}
\end{table}

\begin{table}[]
	\centering
	\caption{Optimum relay location of HD transmission for TSRS.}
	\begin{tabular}{|c|c|c|c|c|c|c|}
		\hline
		\textbf{\begin{tabular}[c]{@{}c@{}}  $\boldsymbol{P_T/\sigma^2}$\\ (dB)\end{tabular}} & $\boldsymbol{a_1}$ & $\boldsymbol{a_3}$  & $(\boldsymbol{\epsilon_{R_l}}, \boldsymbol{\epsilon_{D_{2l}}})$ & $(\boldsymbol{R_{D_l}}, \boldsymbol{R_{D_2}})$  & $(\boldsymbol{x_{R_l}}, \boldsymbol{y_{R_l}})$ & $\boldsymbol{OP_{\min}}$ \\ \hline
		50 & 0.55 & 0.75 & (0, 0) & (0.1, 1) & (0.000, 000) & $4.02\times 10^{-5}$ \\ \hline
		50 & 0.75  & 0.55 & (0, 0) & (0.1, 1) & (0.000, 000) & $8.37\times 10^{-6}$ \\ \hline
		50 & 0.75 & 0.75 & (0, 0) & (0.1, 1)  & (0.000, 000) & $3.39\times 10^{-5}$ \\ \hline
		50 & 0.55  & 0.75 & (0, 0) & (0.1, 1) & (-0.191, 1.959) & $5.89\times 10^{-6}$ \\ \hline
		50 & 0.75 & 0.55 & (0, 0) & (0.1, 1) & (-1.379, 0.276) & $4.91\times 10^{-6}$  \\ \hline
		50  & 0.75 & 0.75 & (0, 0) & (0.1, 1) & (-0.312, 0.468) & $3.13\times 10^{-5}$ \\ \hline
		50 & 0.75 & 0.75 & (0, 0)   & (0.3, 1) & (-1.247, 1.523) & $7.48\times 10^{-5}$ \\ \hline
		50 & 0.75 & 0.75 & (0, 0) & (0.1, 2) & (-0.052, 0.078) & $2.13\times 10^{-1}$ \\ \hline
		55 & 0.75 & 0.75 & (0, 0) & (0.1, 1) & (-1.188, 1.451) & $1.24\times 10^{-8}$  \\ \hline
		50 & 0.75 & 0.75 & $(10^{-4}, 10^{-4})$ & (0.1, 1) & (-0.036, 0.087) & $1.45\times 10^{-1}$ \\ \hline
	\end{tabular}
	\label{RelayLocationResultsHD}
\end{table}

\begin{table}[]
	\centering
	\caption{Optimum distances of HD transmission for TSRS.}
	\begin{tabular}{|c|c|c|c|c|}
		\hline
		$\boldsymbol{d_{S_1R_l}}$ & $\boldsymbol{d_{S_2R_l}}$ & $\boldsymbol{d_{D_1R_l}}$ & $\boldsymbol{d_{D_2R_l}}$ & $\boldsymbol{OP_{\min}}$ \\ \hline
		8.485  & 8.485  & 8.485  & 8.485  & $4.02\times 10^{-5}$ \\ \hline
		8.485  & 8.485  & 8.485  & 8.485  & $8.37\times 10^{-6}$ \\ \hline
		8.485  & 8.485  & 8.485  & 8.485  & $3.39\times 10^{-5}$ \\ \hline
		7.076  & 9.853  & 10.083 & 7.393  & $5.89\times 10^{-6}$ \\ \hline
		7.356  & 7.794  & 9.687  & 9.339 & $4.91\times 10^{-6}$  \\ \hline
		7.935  & 8.613  & 9.037  & 8.393 & $3.13\times 10^{-5}$ \\ \hline
		6.530  & 8.899  & 10.446 & 8.518 & $7.48\times 10^{-5}$ \\ \hline
		8.393  & 8.504  & 8.577  & 8.467 & $2.13\times 10^{-1}$ \\ \hline
		6.622  & 8.870  & 10.353 & 8.507 & $1.24\times 10^{-8}$ \\ \hline
		8.398  & 8.522  & 8.572  & 8.450 & $1.45\times 10^{-1}$ \\ \hline
	\end{tabular}
	\label{HDOptimumDistance}
\end{table}

\section{Conclusion} \label{Conclusion}

A system consisting of two sources and users with FD/HD multi relays, applying NOMA in both sources and relays with up and down link transmissions, is investigated over i.n.i.d. Nakagami-$m$ fading environment for both perfect and imperfect SIC cases, where DF protocol is used. Two different relay selection methods are investigated in accordance with service quality priority, where a high priority is assumed for user 1. Exact and asymptotic OP expressions of both strategies for i.n.i.d. Nakagami-$m$ fade are derived and validated via Monte-Carlo simulation technique. Best relay locations of different schemes including combination of power allocations, data rates, ipSICs, and total powers are illustrated.

\begin{appendices}
\numberwithin{equation}{section}

\section{Outage Probability Derivation of Single Stage Relay Selection }  \label{SSROPDerivation}

  \begin{equation} \label{A1}
  \begin{aligned}
    P_{SSRS}(\gamma_{th_1})&=Pr(\min\{\gamma_{D_1}^{R_{i_{S}}}, \gamma_{D_{1{i_{S}}}}^{D_1}, \gamma_{D_{1{i_{S}}}}^{D_2}\}<\gamma_{th_1})\\
    &=\prod_{i=1}^L Pr(\min\{\gamma_{D_1}^{R_i}, \gamma_{D_{1i}}^{D_1}, \gamma_{D_{1i}}^{D_2}\}<\gamma_{th_1})\\
    &=\prod_{i=1}^L \bigl(1-Pr(\min\{\gamma_{D_1}^{R_i}, \gamma_{D_{1i}}^{D_1}, \gamma_{D_{1i}}^{D_2}\}>\gamma_{th_1})\bigr)\\
        &=\prod_{i=1}^L \bigl(1-Pr(R_i \in K_R)\bigr) ,
      \end{aligned}
  \end{equation}
Due to the independence of channel gains, in turn the independence of  $\gamma_{D_1}^{R_i}$, $\gamma_{D_{1i}}^{D_1}$, and $\gamma_{D_{1i}}^{D_2}$, $Pr(R_i \in K_R)=Pr(\gamma_{D_1}^{R_i}>\gamma_{th_1})Pr( \gamma_{D_{1i}}^{D_1}>\gamma_{th_1})Pr(\gamma_{D_{1i}}^{D_2}>\gamma_{th_1})$. Therefore, in order to obtain a closed form of $ P_{SSRS}(\gamma_{th_1})$, it is sufficient to derive CDFs of $\gamma_{D_1}^{R_i}$, $\gamma_{D_{1i}}^{D_1}$, and $\gamma_{D_{1i}}^{D_2}$:

  \begin{equation} \label{A2}
  \begin{aligned}
   F_{\gamma_{D_1}^{R_i}}(\gamma_{th_1})&=Pr(\gamma_{D_1}^{R_i}<\gamma_{th_1})\\
    &=Pr(\frac{a_1 \rho_{S_{1i}}|g_{S_i}|^2}{a_2\rho_{S_{1i}}|h_{S_i}|^2+\varpi_i\rho_{R_i}|h_{R_iR_i}|^2+1}<\gamma_{th_1}) \\
    &=Pr \biggl(|g_{S_i}|^2< \frac{\gamma_{th_1}}{a_1 \rho_{S_{1i}}} \biggl(a_2\rho_{S_{1i}}|h_{S_i}|^2+\varpi_i\rho_{R_i}|h_{R_iR_i}|^2+1\biggr)\biggr)\\
    &=\int_{0}^{\infty} \int_{0}^{\infty} F_{|g_{S_i}|^2}\biggl( \frac{\gamma_{th_1}}{a_1 \rho_{S_{1i}}} \bigl(a_2\rho_{S_{1i}}x+\varpi_i\rho_{R_i}y+1\bigr)\biggr) f_{|h_{S_i}|^2}(x)f_{|h_{R_iR_i}|^2}(y)dxdy,
      \end{aligned}
  \end{equation}
where the CDFs and PDFs of the square of all channel gains are given in (\ref{A3}).
  \begin{equation} \label{A3}
  \begin{aligned}
   F_{|g_{S_i}|^2}(x)&=\frac{\gamma\biggl(m_{S_{1i}}, m_{S_{1i}}x/\Omega_{S_{1i}}\biggr)}{\Gamma(m_{S_{1i}})}\\
   f_{|g_{S_i}|^2}(x)&=\frac{m_{S_{1i}}^{m_{S_{1i}}}x^{m_{S_{1i}}-1} e^{-m_{S_{1i}}x/\Omega_{S_{1i}}} }{\Omega_{S_{1i}}^{m_{S_{1i}}}\Gamma(m_{S_{1i}})}\\
   F_{|h_{S_i}|^2}(x)&=\frac{\gamma\biggl(m_{S_{2i}}, m_{S_{2i}}x/\Omega_{S_{2i}}\biggr)}{\Gamma(m_{S_{2i}})}\\
   f_{|h_{S_i}|^2}(x)&=\frac{m_{S_{2i}}^{m_{S_{2i}}}x^{m_{S_{2i}}-1} e^{-m_{S_{2i}}x/\Omega_{S_{2i}}} }{\Omega_{S_{2i}}^{m_{S_{2i}}}\Gamma(m_{S_{2i}})}\\
   F_{|g_{R_{i1}}|^2}(x)&=\frac{\gamma\biggl(m_{D_{1i}}, m_{D_{1i}}x/\Omega_{D_{1i}}\biggr)}{\Gamma(m_{D_{1i}})}\\
   f_{|g_{R_{i1}}|^2}(x)&=\frac{m_{D_{1i}}^{m_{D_{1i}}}x^{m_{D_{1i}}-1} e^{-m_{D_{1i}}x/\Omega_{D_{1i}}} }{\Omega_{D_{1i}}^{m_{D_{1i}}}\Gamma(m_{D_{1i}})}\\  
   F_{|h_{R_{i2}}|^2}(x)&=\frac{\gamma\biggl(m_{D_{2i}}, m_{D_{2i}}x/\Omega_{D_{2i}}\biggr)}{\Gamma(m_{D_{2i}})}\\
   f_{|h_{R_{i2}}|^2}(x)&=\frac{m_{D_{2i}}^{m_{D_{2i}}}x^{m_{D_{2i}}-1} e^{-m_{D_{2i}}x/\Omega_{D_{2i}}} }{\Omega_{D_{2i}}^{m_{D_{2i}}}\Gamma(m_{D_{2i}})}\\ 
 \end{aligned}
  \end{equation}
Similarly, CDFs and PDFs of ipSIC and SI are given in (\ref{A4})
  \begin{equation} \label{A4}
  \begin{aligned}
   F_{|\tilde{g}_{S_i}|^2}(x)&=\frac{\gamma\biggl(\tilde{m}_{R_i}, \tilde{m}_{R_i}x\biggr)}{\Gamma(\tilde{m}_{R_i})}\\
   f_{|\tilde{g}_{S_i}|^2}(x)&=\frac{\tilde{m}_{R_i}^{\tilde{m}_{R_i}}x^{\tilde{m}_{R_i}-1} e^{-\tilde{m}_{R_i}x} }{\Gamma(\tilde{m}_{S_{1i}})}\\
   F_{|\tilde{h}_{R_{i2}}|^2}(x)&=\frac{\gamma\biggl(\tilde{m}_{D_{2i}}, \tilde{m}_{D_{2i}}x\biggr)}{\Gamma(\tilde{m}_{D_{2i}})}\\
   f_{|\tilde{h}_{R_{i2}}|^2}(x)&=\frac{\tilde{m}_{D_{2i}}^{\tilde{m}_{D_{2i}}}x^{\tilde{m}_{D_{2i}}-1} e^{-\tilde{m}_{D_{2i}}x}}{\Gamma(\tilde{m}_{D_{2i}})}\\ 
   F_{|h_{R_iR_i}|^2}(x)&=\frac{\gamma\biggl(m_{R_iR_i}, m_{R_iR_i}x/\Omega_{R_iR_i}\biggr)}{\Gamma(m_{R_iR_i})}\\
   f_{|h_{R_iR_i}|^2}(x)&=\frac{m_{R_iR_i}^{m_{R_iR_i}}x^{m_{R_iR_i}-1} e^{-m_{R_iR_i}x/\Omega_{R_iR_i}} }{\Omega_{R_iR_i}^{m_{R_iR_i}}\Gamma(m_{R_iR_i})}\\   
 \end{aligned}
  \end{equation}
After substitution of $ F_{|g_{S_i}|^2}(x)$, $ f_{|h_{S_i}|^2}(x)$, and $ f_{|h_{R_iR_i}|^2}(x)$  in (\ref{A2}) and carrying out the following steps, the closed form of  $ F_{\gamma_{D_1}^{R_i}}(\gamma_{th_1})$ is derived as in (\ref{A5}):
\begin{itemize}
\item [$\bullet$]  Firstly, use the equality for upper incomplete gamma function given in \cite[{eq. (8.352)}]{Ryzhik-2007}:
$\gamma\bigl(m_{S_{1i}}, z\bigr)=(m_{S_{1i}}-1)!\biggl[1- e^{-z}\sum_{p_1=0}^{m_{S_{1i}}-1} \frac{z^{p_1}}{p1!}\biggr]$.
\item [$\bullet$]  Secondly, apply Binomial theorem to the term $\bigl(a_2\rho_{S_{1i}}x+\varpi_i\rho_{R_i}y+1\bigr)^{p_1}$ twice to obtain $\bigl(a_2\rho_{S_{1i}}x+\varpi_i\rho_{R_i}y+1\bigr)^{p_1}=\sum_{p_2=0}^{p_1}\sum_{p_3=0}^{p_2}\binom{p_1}{p_2} \binom{p_2}{p_3}\bigl(a_2\rho_{S_{1i}}x\bigr)^{p_3}\bigl(\varpi_i\rho_{R_i}y\bigr)^{p_2-p_3}$.
\item [$\bullet$] Finally, after proper rearrangement, obtain two integrals which are in the form of Gamma function given in \cite[{eq. (8.310)}]{Ryzhik-2007}, namely,  $\Gamma(z)=\int_{0}^{\infty} e^{-t}t^{z-1}dt$.
\end{itemize} 
 \begin{equation} \label{A5}
 \begin{aligned}
    F_{\gamma_{D_1}^{R_i}}(\gamma_{th_1})=&1- e^{-\frac{m_{S_{1i}}\gamma_{th_1}}{a_1 \rho_{S_{1i}}\Omega_{S_{1i}} }} \sum_{p_1=0}^{m_{S_{1i}}-1} \sum_{p_2=0}^{p_1} \sum_{p_3=0}^{p_2} \biggl[ \biggl\{    \biggl(\frac{m_{S_{1i}}\gamma_{th_1}}{a_1\Omega_{S_{1i}}}\biggr ) ^{p_1}  \biggl(\frac{a_2\Omega_{S_{2i}}}{ m_{S_{2i}}}\biggr )^{p_3} \binom{p_1}{p_2}\binom{p_2}{p_3} \\ 
    &\Gamma(p_3+m_{S_{2i}}) \Gamma(p_2-p_3+m_{R_i}) \biggl( \frac{\varpi_i\rho_{R_i}\Omega_{R_iR_i}}{m_{R_i}}  \biggr)^{p_2-p_3} \biggr\} \bigg/ \biggl\{ \Gamma(m_{S_{2i}}) \Gamma(m_{R_i}) \\ 
    &\Gamma(p_1+1)    \biggl(1+ \frac{m_{S_{1i}}\gamma_{th_1}a_2\Omega_{S_{2i}}}{m_{S_{2i}}a_1m_{S_{1i}} }  \biggr)^{m_{S_{2i}}+p_3}    \biggl(1+ \frac{m_{S_{1i}}\gamma_{th_1}\varpi_i \rho_{R_i} \Omega_{R_iR_i}}{m_{R_i}a_1  \rho_{S_i}\Omega_{S_{1i}} }  \biggr)^{m_{R_i}+p_2-p_3}\\ &(\rho_{S_i})^{p_1-p_3}      \biggr\} \biggr]
 \end{aligned}.
 \end{equation}

The closed form of $ F_{\gamma_{D_{1i}}^{D_1}}(\gamma_{th_1})$ is 
  \begin{equation} \label{A6}
  \begin{aligned}
   F_{\gamma_{D_{1i}}^{D_1}}(\gamma_{th_1})&=Pr(\gamma_{D_{1i}}^{D_1}<\gamma_{th_1})\\
    &=Pr(\frac{a_3\rho_{D_{1i}}|g_{R_{i1}}|^2}{a_4\rho_{D_{1i}}|g_{R_{i1}}|^2+1}<\gamma_{th_1}) \\
    &
    =\left\{
    \begin{array}{ll}
          1 & \gamma_{th_1}\ge \frac{a_3}{a_4} \\
        Pr \bigl(|g_{R_{i1}}|^2< \frac{\gamma_{th_1}}{a_3 \rho_{D_{1i}}-a_4 \rho_{D_{1i}}\gamma_{th_1}}\bigr) & \gamma_{th_1} < \frac{a_3}{a_4}\\
    \end{array} 
    \right.\\
    &=\left\{
        \begin{array}{ll}
              1 & \gamma_{th_1}\ge \frac{a_3}{a_4} \\
            \frac{ \gamma\biggl(m_{D_{1i}}, \frac{m_{D_{1i}}\gamma_{th_1}}{\rho_{D_{1i}}\Omega_{D_{1i}}( a_3-a_4 \gamma_{th_1})}  \biggr) }{\Gamma(m_{D_{1i}})} & \gamma_{th_1} < \frac{a_3}{a_4}\\
        \end{array} 
        \right.\\
      \end{aligned}.
  \end{equation}
In the same manner,  $F_{\gamma_{D_{1i}}^{D_2}}(\gamma_{th_1})$ is
\begin{equation} \label{A7}
\begin{aligned}
F_{\gamma_{D_{1i}}^{D_2}}(\gamma_{th_1})&=Pr(\gamma_{D_{1i}}^{D_2}<\gamma_{th_1})\\
&=\left\{
\begin{array}{ll}
1 & \gamma_{th_1}\ge \frac{a_3}{a_4} \\
\frac{ \gamma\biggl(m_{D_{2i}}, \frac{m_{D_{2i}}\gamma_{th_1}}{\rho_{D_{2i}}\Omega_{D_{2i}}( a_3-a_4 \gamma_{th_1})}  \biggr) }{\Gamma(m_{D_{2i}})} & \gamma_{th_1} < \frac{a_3}{a_4}\\
\end{array} 
\right.\\
\end{aligned}.
\end{equation}
Inserting complementaries of (\ref{A5}), (\ref{A6}), and (\ref{A7}) into (\ref{A1}) together with the fact that $\Gamma(m)=\gamma(m,x)+\Gamma(m,x)$, the closed form of  $P_{SSRS}(\gamma_{th_1})$ reduces to (\ref{Denklem_12}).

\section{Outage Probability Derivation of Two Stage Relay Selection } 

$ P_{TSRS}(\gamma_{th_1},\gamma_{th_2})$ is rearranged as 
\begin{equation} \label{B1}
\begin{aligned}
 P_{TSRS}(\gamma_{th_1},\gamma_{th_2})=&\sum_{l=0}^L Pr(\min(\gamma_{D_{2i_{T}}}^{D_2},\gamma_{D_2}^{R_{i_{T}}})< \gamma_{th_2}, |K_R|=l)\\
 =&\sum_{l=0}^L Pr(\min(\gamma_{D_{2i_{T}}}^{D_2},\gamma_{D_2}^{R_{i_{T}}})< \gamma_{th_2} \big/ |K_R|=l)Pr(|K_R|=l)\\
 =&\sum_{l=0}^L   \prod_{\substack{j=1 \\ R_i \in K_R}}^{l} \bigl(1-Pr\bigl(\min(\gamma_{D_{2i}}^{D_2},\gamma_{D_2}^{R_i})>\gamma_{th_2} \big/ R_i\in K_R \bigr)\bigr)Pr(|K_R|=l)\\
  =&\sum_{l=0}^L  \binom{L}{l}  \prod_{\substack{j=1 \\ R_i \in K_R}}^{l} \bigl( 1-P_{\phi}(\gamma_{th_1},\gamma_{th_2}) \bigr)
  \prod_{\substack{j=1 \\ R_i \notin K_R}}^{L-l} \bigl(1- Pr(R_i\in K_R) \bigr)\\ 
  &\times \prod_{\substack{j=1 \\ R_i \in K_R}}^{l}  Pr(R_i\in K_R) 
  \end{aligned}\\.
\end{equation}
The closed form expression of $Pr(R_i\in K_R)$ in (\ref{B1}) is obtained in the derivation of $P_{SSRS}(\gamma_{th_1})$  in (\ref{Denklem_13})
and $P_{\phi}(\gamma_{th_1},\gamma_{th_2})= Pr(\min(\gamma_{D_{2i}}^{D_2},\gamma_{D_2}^{R_i})> \gamma_{th_2} / \min\{\gamma_{D_1}^{R_i}, \gamma_{D_{1i}}^{D_1}, \gamma_{D_{1i}}^{D_2}\}> \gamma_{th_1})$, by the aid of Bayes theorem, we get 
\begin{equation} \label{B3}
\begin{array} {ll}
P_{\phi}(\gamma_{th_1},\gamma_{th_2})&=\frac{Pr(\min(\gamma_{D_{2i}}^{D_2},\gamma_{D_2}^{R_i})> \gamma_{th_2}, \min(\gamma_{D_1}^{R_i}, \gamma_{D_{1i}}^{D_1}, \gamma_{D_{1i}}^{D_2})> \gamma_{th_1})}{P_{R_i \in K_R}(\gamma_{th_1})} \\
&=\frac{P_{\phi_1}(\gamma_{th_1},\gamma_{th_2})P_{\phi_2}(\gamma_{th_1},\gamma_{th_2})P_{\phi_3}(\gamma_{th_1},\gamma_{th_2})}{P_{R_i \in K_R}(\gamma_{th_1})}
\end{array},
\end{equation}
where $Pr(\min(\gamma_{D_{2i}}^{D_2},\gamma_{D_2}^{R_i})> \gamma_{th_2}, \min(\gamma_{D_1}^{R_i}, \gamma_{D_{1i}}^{D_1}, \gamma_{D_{1i}}^{D_2})> \gamma_{th_1})=Pr\bigl(\gamma_{D_{2i}}^{D_2} > \gamma_{th_2}, \gamma_{D_2}^{R_i} > \gamma_{th_2}, \gamma_{D_1}^{R_i} > \gamma_{th_1}, \gamma_{D_{1i}}^{D_1} > \gamma_{th_1}, \gamma_{D_{1i}}^{D_2}> \gamma_{th_1} \bigr)$. Using the independence of the channels, it can be rephrased into  three independent components as in the  numerator of the second line of (\ref{B3}). These components, namely,  $P_{\phi_1}(\gamma_{th_1},\gamma_{th_2})$, $P_{\phi_2}(\gamma_{th_1},\gamma_{th_2})$ and $P_{\phi_3}(\gamma_{th_1},\gamma_{th_2})$, are defined as follows:
\begin{equation} \label{B4}
\begin{array}{lll}
&\begin{split}
P_{\phi_1}(\gamma_{th_1},\gamma_{th_2})&=Pr\biggl( |\tilde{h}_{R_{i2}}|^2 >  \frac{a_4\rho_{D_{2i}}|h_{R_{i2}}|^2-\gamma_{th_2}}{\gamma_{th_2}a_3 \epsilon_{D_{2i}}\rho_{D_{2i}}} ,  |h_{R_{i2}}|^2 > \frac{\gamma_{th_1}}{a_3\rho_{D_{2i}}-a_4\rho_{D_{2i}}\gamma_{th_1}}  \biggr)\\
&=Pr\biggl( |\tilde{h}_{R_{i2}}|^2 <   \frac{a_4\rho_{D_{2i}}|h_{R_{i2}}|^2-\gamma_{th_2}}{\gamma_{th_2}a_3 \epsilon_{D_{2i}}\rho_{D_{2i}}} ,  |h_{R_{i2}}|^2 > U_{max} \biggr)
\end{split}\\
&\begin{aligned}
P_{\phi_2}(\gamma_{th_1},\gamma_{th_2})=Pr\biggl(|g_{R_{i1}}|^2 >  \frac{\gamma_{th_1}}{a_3\rho_{D_{1i}}-a_4\rho_{D_{1i}}\gamma_{th_1}}  \biggr)
\end{aligned}\\
&\begin{split}
P_{\phi_3}(\gamma_{th_1},\gamma_{th_2})=&Pr\biggl(|h_{S_i}|^2 >  \frac{\gamma_{th_2}\bigl(\varpi_i \rho_{R_i} |h_{R_iR_i}|^2+1\bigr)}{a_2\rho_{S_i}},
\\
 &|g_{S_i}|^2 >  \frac{\gamma_{th_1}\bigl(a_2 \rho_{S_i} |h_{S_i}|^2 +  \varpi_i \rho_{R_i} |h_{R_iR_i}|^2+1\bigr)}{a_1\rho_{S_i}} \biggr)
\end{split}
\end{array}.
\end{equation}
where $U_{max}=\max\bigl( \frac{\gamma_{th_2}}{a_4\rho_{D_{2i}}},\frac{\gamma_{th_1}}{ \rho_{D_{2i}}(a_3-a_4\gamma_{th_1})} \bigr)$.
To find closed form of $P_{\phi_1}(\gamma_{th_1},\gamma_{th_2})$, we need to take the integral given in (\ref{B5}):
\begin{equation} \label{B5}
\begin{split}
P_{\phi_1}(\gamma_{th_1},\gamma_{th_2})&=Pr\biggl( |\tilde{h}_{R_{i2}}|^2 <   \frac{a_4\rho_{D_{2i}}|h_{R_{i2}}|^2-\gamma_{th_2}}{\gamma_{th_2}a_3 \epsilon_{D_{2i}}\rho_{D_{2i}}} ,  |h_{R_{i2}}|^2 > U_{max} \biggr)\\
&=\int_{U_max}^{\infty}F_{|\tilde{h}_{R_{i2}}|^2}\biggl(\frac{a_4\rho_{D_{2i}}x-\gamma_{th_2}}{\gamma_{th_2}a_3 \epsilon_{D_{2i}}\rho_{D_{2i}}}\biggr)f_{|h_{R_{i2}}|^2}(x)dx.
\end{split}
\end{equation}
and it is attained as in (\ref{Denklem_17}) after implementation of the following steps:
\begin{itemize}
\item [$\bullet$] Firstly, insert equivalences of $F_{|\tilde{h}_{R_{i2}}|^2}(x)$ and $f_{|h_{R_{i2}}|^2}(x)$ from (\ref{A4}) and (\ref{A3}) into (\ref{B5}). 
\item [$\bullet$]  Secondly, insert the identity \cite[{eq. (8.352.1)}]{Ryzhik-2007} for lower incomplete Gamma function   into (\ref{B5}) to obtain $\gamma\bigl(\tilde{m}_{D_{2i}}, z\bigr)=(\tilde{m}_{D_{2i}}-1)!\biggl[1- e^{-z}\sum_{p_1=0}^{\tilde{m}_{D_{2i}}-1} \frac{z^{p_1}}{p1!}\biggr]$. 
\item [$\bullet$]  Thirdly, use Binomial theorem to get $\bigl(\frac{a_4\rho_{D_{2i}}}{\gamma_{th_2}}x-1\bigr)^{p_1}=\sum_{p_2=0}^{p_1}\binom{p_1}{p_2}(-1)^{p_1-p_2}\bigl(\frac{a_4\rho_{D_{2i}}}{\gamma_{th_2}}x\bigr)^{p_2}$. 
\item [$\bullet$] Finally, after proper rearrangements, use the identity \cite[{eq. (8.350.2)}]{Ryzhik-2007} for upper incomplete Gamma function, namely,  $\Gamma(z, x)=\int_{x}^{\infty} e^{-t}t^{z-1}dt$ to get closed form of the integral in (\ref{B5}), thereafter, closed form of  $P_{\phi_1}(\gamma_{th_1},\gamma_{th_2})$ is reached as given in (\ref{Denklem_17}).
\end{itemize} 

The second probability, $P_{\phi_2}(\gamma_{th_1},\gamma_{th_2})$, is 
\begin{equation} \label{B6}
\begin{aligned}
P_{\phi_2}(\gamma_{th_1},\gamma_{th_2})=\left\{
\begin{array}{ll}
0 & \gamma_{th_1}\ge \frac{a_3}{a_4} \\
1-\frac{ \gamma\biggl(m_{D_{1i}},  \frac{m_{D_{1i}}\gamma_{th_1}}{\rho_{D_{1i}} \Omega_{D_{1i}}(a_3-\gamma_{th_1}a_4)}  \biggr)}{\Gamma(m_{D_{1i}}) } & \gamma_{th_1} < \frac{a_3}{a_4}\\
\end{array} 
\right.\\
\end{aligned}
\end{equation}
and the third part of $P_{\phi}(\gamma_{th_1},\gamma_{th_2})$ can be derived as
\begin{equation} \label{B7}
\begin{array}{lll}
&\begin{split}
P_{\phi_3}(\gamma_{th_1},\gamma_{th_2})=&\int_{0}^{\infty}\int_{0}^{\infty} \int_{\frac{\gamma_{th_2}\bigl( a_1\epsilon_{R_i}\rho{S_i}\psi+\varpi_i\rho{R_i}z+1 \bigr)}{a_2\rho{S_i}}}^{\infty} \int_{\frac{\gamma_{th_1}\bigl(a_2\rho_{S_i}y+\varpi_i\rho{R_i}z+1 \bigr)}{a_1\rho{S_i}}}^{\infty} f_{|g_{S_i}|^2}(x) \\
 & f_{|h_{S_i}|^2}(y) f_{|h_{R_iR_i}|^2}(z)  f_{|\tilde{g}_{S_i}|^2}(\psi)dxdydzd\psi
\end{split}
\end{array}.
\end{equation}
The closed form of $P_{\phi_3}(\gamma_{th_1},\gamma_{th_2})$ can be derived after carrying out the steps provided below:
\begin{itemize}
\item [$\bullet$] The order of integration is respectively with respect to $x$, $y$, $z$, and $\psi$. 
\item [$\bullet$] Substitute equivalences of $f_{|g_{S_i}|^2}(x)$, $f_{|h_{S_i}|^2}(y)$, $f_{|h_{R_iR_i}|^2}(z)$, and $f_{|\tilde{g}_{S_i}|^2}(\psi)$ into (\ref{B7}) from (\ref{A3}) and (\ref{A4}). Then make change of variable for $x$ to convert the integral into the form of upper incomplete Gamma function \cite[{eq. (8.350.2)}]{Ryzhik-2007}. 

\item [$\bullet$] Use the identity given in \cite[{eq. (8.350.2)}]{Ryzhik-2007} for upper incomplete Gamma function to obtain $\Gamma(m_{S_{1i}},x)=\Gamma(m_{S_{1i}})e^{-x}\sum_{q_1=0}^{m_{S_{1i}}-1}\frac{x^{q_1}}{\Gamma(q_1+1)}$.

\item [$\bullet$] Apply Binomial theorem twice to derive $(a_2\rho_{S_i}y+\varpi_i\rho_{R_i}z+1)^{q_1}=\sum_{q_2=0}^{q_1}\sum_{q_3=0}^{q_2}\\\binom{q_1}{q_2}\binom{q_2}{q_3}(a_2\rho_{S_i}y)^{q_2}(\varpi_i\rho_{R_i}z)^{q_2-q_3}$ and represent the triple summations as $\\\sum_{q_1=0}^{m_{S_{1i}}-1}\sum_{q_2=0}^{q_1}\sum_{q_3=0}^{q_2}=\sum_{(q_1, q_2,q_3)=\overrightarrow{\rm \boldsymbol{0}} }^{(m_{S_{1i}}-1,q_1,q_2)}$, where $\overrightarrow{\rm \boldsymbol{0}}$ is a three dimensional zero vector.

\item [$\bullet$] Integration over $y$ is similar to integration over $x$. Therefore, track the same steps: Use expansion of upper incomplete Gamma function for integer numbers to get $\Gamma(q_3+m_{S_{2i}},x)=\Gamma(q_3+m_{S_{2i}})e^{-x}\sum_{k_1=0}^{q_3+m_{S_{2i}}-1}\frac{x^{k_1}}{\Gamma(k_1+1)}$. Then, apply Binomial theorem twice to derive $(a_1 \epsilon_{R_i}\rho_{S_i}\psi+\varpi_i\rho_{R_i}z+1)^{k_1}=\sum_{k_2=0}^{k_1}\sum_{k_3=0}^{k_2}\binom{k_1}{k_2}\binom{k_2}{k_3}(a_1 \epsilon_{R_i}\rho_{S_i}\psi)^{k_2}(\varpi_i\rho_{R_i}z)^{k_2-k_3}$ and represent the triple summations as $\sum_{k_1=0}^{q_3+m_{S_{2i}}-1}\sum_{k_2=0}^{k_1}\sum_{k_3=0}^{k_2}=\sum_{(k_1, k_2,k_3)=\overrightarrow{\rm \boldsymbol{0}} }^{(q_3+m_{S_{2i}}-1,k_1,k_2)}$.
\item [$\bullet$]  After proper simplifications, remaining two integrals are independent of each other and in the form of Gamma function given in  \cite[{eq. (8.310)}]{Ryzhik-2007} which completes closed form derivation of $P_{\phi_3}(\gamma_{th_1},\gamma_{th_2})$ provided in (\ref{Denklem_19}).
\end{itemize}

\end{appendices}


\end{document}